\documentclass[%
 reprint,longbibliography,
showpacs,preprintnumbers,
bibnotes,
 amsmath,amssymb,
prb,
]{revtex4-2}
\usepackage{graphicx}
\usepackage[english]{babel}
\usepackage{microtype}       
\usepackage[utf8]{inputenc}
\usepackage[breaklinks=true,colorlinks=true,linkcolor=blue,urlcolor=blue,citecolor=blue]{hyperref}

\usepackage[dvipsnames]{xcolor}      
\usepackage{mathtools}
\mathtoolsset{showonlyrefs=true}
\usepackage{braket}
\usepackage{tikz}

 \usepackage{longtable}
  \usepackage{booktabs}
 \usepackage{siunitx}
 \usepackage{csquotes}
 
 \usepackage{multirow}
\usepackage{amssymb}

\newcommand{\eqrefs}[2]{Eqs.~\eqref{#1} and \eqref{#2}}

\newcommand{\tn}[1]{\textnormal{#1}}

\newcommand{\fluc}{\delta\hat{G}}
\newcommand{\hh}{\mathrm{H}}
\newcommand{\HF}{\mathrm{HF}}
\newcommand{\cfluc}{\Delta G^\lambda}
\newcommand{\CC}{\mathbb{C}}

\newcommand{\chiR}{\chi^\mathrm{R}}

\usepackage{dcolumn}
\usepackage{bm}

\makeatletter
\g@addto@macro\bfseries{\boldmath}
\newcommand*{\balancecolsandclearpage}{%
   \close@column@grid
   \clearpage
   \twocolumngrid
 }
\makeatother

\begin{document}

\title{Quantum Fluctuations Approach to the Nonequilibrium $GW$-Approximation II: Density Correlations and Dynamic Structure Factor}

\author{Erik Schroedter, Björn Jakob Wurst, Jan-Philip Joost, and
Michael Bonitz
 \email{bonitz@theo-physik.uni-kiel.de}}
\affiliation{
Institut f\"ur Theoretische Physik und Astrophysik, 
Christian-Albrechts-Universit\"{a}t zu Kiel, D-24098 Kiel, Germany\\ and
Kiel Nano, Surface and Interface Science KiNSIS, Kiel University, Germany
}

\date{\today}

\begin{abstract}
The quantum dynamics of correlated fermionic or bosonic many-body systems following external excitation can be successfully studied using nonequilibrium Green functions (NEGF) or reduced density matrix methods. Approximations are introduced via a proper choice of the many-particle selfenergy or decoupling of the BBGKY-hierarchy, respectively. These approximations are based on Feynman's diagram approaches or on cluster expansions into single-particle and correlation operators.
In a recent paper [E. Schroedter, J.-P. Joost, and M. Bonitz, Cond. Matt. Phys. \textbf{25}, 23401 (2022)] we have presented a different approach where, instead of equations of motion for the many-particle NEGF (or density operators), equations for the correlation functions of fluctuations are analyzed. In particular, we derived the stochastic $GW$ and polarization approximations that are closely related to the nonequilibrium $GW$ approximation.
Here, we extend this approach to the computation of two-time observables depending on the specific ordering of the underlying operators. In particular, we apply this extension to the calculation of  the density correlation function and dynamic structure factor of correlated Hubbard clusters in and out of equilbrium.
\end{abstract}

\maketitle

\section{Introduction}\label{s:intro}
The dynamics of quantum many-body systems following external excitation are of high interest in many areas such as dense plasmas, nuclear matter, ultracold atoms or correlated solids. There is a large variety of methods available to simulate such systems which include real-time quantum Monte Carlo, density matrix renormalization group approaches, time-dependent density functional theory and quantum kinetic theory. Among the many-particle observables that are accessible in experiments, a central role is played by the correlation functions of density or spin fluctuations and the corresponding dynamic structure factors, see e.g. Ref. \cite{giuliani_vignale_2005} for an overview. To compute these quantities with correlation effects taken into account there exist a variety of equilibrium simulations. The most accurate results have been obtained from quantum Monte Carlo simulations for correlated solids, e.g. \cite{moreo_prb_93,lee_prb_03,assaad_prb_06}, as well as warm dense matter \cite{dornheim_prl_18,dornheim_pop_23,bonitz_pop_20} where also the non-linear response has been analyzed \cite{bonitz-etal.94pre,dornheim_prl_20, dornheim_prr_21}.
In addition, there exists a variety of nonequilibrium approaches, including dynamical mean field theory, e.g. \cite{gull_prb_19,gull_prr_20}, time-dependent DMRG, e.g. \cite{pereira_prb_12}, and nonequilibrium Green functions (NEGF), cf.  \cite{kwong_prl_00} and references therein.

Here we concentrate on the NEGF approach  \cite{keldysh64, stefanucci-book,balzer-book} 
because it can rigorously describe the quantum dynamics of correlated systems in more than one dimension, e.g. \cite{schluenzen_prb16}. However, NEGF simulations are computationally expensive, primarily, due to their cubic scaling with the simulation time $N_t$ (number of time steps). Only recently,
linear scaling with $N_t$ could be achieved within the G1--G2 scheme \cite{schluenzen_prl_20,joost_prb_20} which could be demonstrated even for advanced selfenergies, including the $GW$ and the $T$-matrix approximations. Even the nonequilibrium dynamically screened ladder approximation, which selfconsistently combines dynamical screening and strong coupling, is now feasible, at least for lattice models \cite{joost_prb_22,donsa_prb_23}. 

The advantage of time linear scaling of the G1--G2 scheme comes at a price: the simultaneous propagation of the time-diagonal single-particle and correlated two-particle Green functions, $G_1(t)$ and $\mathcal{G}_2(t)$, requires a large computational effort for computing and storing all matrix elements of $\mathcal{G}_2$. For example, the CPU time of GW-G1-G2 simulations scales as $N_b^6$, where $N_b$ is the basis dimension. Even though this difficulty can be relieved using massively parallel computer hardware or embedding selfenergy approaches \cite{stefanucci-book,balzer_prb_23}, it is well worth to look for alternative formulations of the problem that are more suitable for computations, ideally without loss of accuracy.

In Ref.~\cite{schroedter_cmp_22} an alternative formulation of the quantum many-body problem was presented that is based on a stochastic approach to the dynamics of quantum fluctuations. Extending earlier stochastic concepts in the kinetic theory of classical systems, due to Klimontovich, e.g. \cite{klimontovich_jetp_57, klimontovich_jetp_72, klimontovich_1982}, and quantum systems by Ayik, Lacroix~\cite{ayik_plb_08, lacroix_prb14,lacroix_epj_14} and many others, e.g. \cite{filinov_prb_2, polkovnikov_ap_10}, we derived an equation of motion for the single-particle fluctuations, $\delta \hat G$ [see Eq.~\eqref{eq:deltag-def} below], that is equivalent to the nonequilibrium $GW$ approximation, in the weak coupling limit. 

Here we extend the results of Ref.~\cite{schroedter_cmp_22} to the nonequilibrium dynamics of two-time quantities (and their Fourier transform) such as the density correlation function (and the dynamic structure factor). This first requires to obtain a semi-classical approach to the computation of commutators of operators. This is achieved within a multiple ensembles (ME) approach. With this extension we are then able to compute the density response function and dynamic structure factor, both, in the ground state and for a far from equilibrium situation following an external excitation. This constitutes a significant extension of the quantum fluctuations approach of Ref.~\cite{schroedter_cmp_22} that is applicable to large systems and long simulation times.

This paper is structured as follows. In Sec.~\ref{s:quantum_fluctuations_approach} we introduce the quantum fluctuations approach and establish its connection to the exchange-correlation function of NEGF theory. Here we also derive the expressions for the dynamic structure factor and the density response function,
This is followed, in Sec.~\ref{s:smf_theory},  by an introduction to our stochastic approach to quantum fluctuations. Then, in Sec.~\ref{s:numerics} we present our numerical results for small and moderate size Hubbard clusters. A summary and outlook is given in Sec.~\ref{s:discussion}.

\section{Quantum Fluctuations Approach} \label{s:quantum_fluctuations_approach}

\subsection{Notation and Definitions} \label{ss:notation_and_definitions}
In the following, we use the formalism of second quantization, which is characterized by the bosonic/fermionic creation $(\hat{c}^\dagger_i)$ and annihilation $(\hat{c}_i)$ operators and the respective single-particle basis of the underlying single-particle Hilbert space $\mathcal{H}$, which induces the so-called Fock space $\mathcal{F}$. These operators have the following properties:
\begin{align}
        \Big[\hat{c}_i,\hat{c}_j^\dagger\Big]_{\mp}=\delta_{ij},\quad \Big[\hat{c}_i,\hat{c}_j\Big]_{\mp}=\Big[\hat{c}^\dagger_i,\hat{c}^\dagger_j\Big]_{\mp}=0\,, \label{eq:ladder_operators}
\end{align}
where the upper/lower sign refers to bosons/fermions respectively. Here, we consider a quantum many-particle system, which is described by a generic Hamiltonian of the form 
\begin{equation}
    \hat{H}(t)= \sum_{ij}h_{ij}(t)\hat{c}^\dagger_i\hat{c}_j+\frac{1}{2}\sum_{ijkl}w_{ijkl}(t)\hat{c}_i^\dagger\hat{c}^\dagger_j\hat{c}_l\hat{c}_k\,, \label{eq:generic_Hamiltonian}
\end{equation}
where $h$ denotes the single-particle contributions (from the kinetic energy and an external potential), and a general pair interaction $w$. Notice that both, $h$ and $w$, are allowed to be time-dependent in order to account for changes in the external potential, e.g., due to lasers \cite{kwong-etal.98pss},  particle impact \cite{balzer_prb16, balzer_prl_18,niggas_prl_22}, or a change of the confinement potential \cite{schluenzen_prb16}, whereas the time dependence of the interaction potential allows for the computation of a correlated initial state from an uncorrelated state via the adiabatic switching method. Additionally, the interaction tensor $w$ obeys the symmetries 
\begin{align}
    w_{ijkl}(t)&=w_{jilk}(t)=[w_{klij}(t)]^*\,. \label{eq:w_symmetries}
\end{align}
The central quantity of the NEGF theory is the one-body Green function which is defined on the Keldysh contour $\mathcal{C}$ for contour-time arguments $z$ and $z'$ as 
\begin{align}
    G_{ij}(z,z')\coloneqq \frac{1}{\mathrm{i}\hbar}\left\langle\mathcal{T}_\mathcal{C}\left\{\hat{c}_i(z)\hat{c}^\dagger_j(z')\right\}\right\rangle\,,
\end{align}
where $\mathcal{T}_\mathcal{C}$ denotes the time-ordering operator on the contour. Averaging is performed with the correlated unperturbed density operator of the system. In the following, it will be sufficient to consider the correlation functions $G^\gtrless$ for real time arguments. We define these functions and the corresponding operators as 
\begin{align}
     G^\gtrless_{ij}(t,t')&\coloneqq 
     \left\langle\hat{G}^\gtrless_{ij}(t,t')\right\rangle\,, \\     
     \hat{G}^<_{ij}(t,t')&\coloneqq \pm\frac{1}{\mathrm{i}\hbar}\hat{c}^\dagger_j(t')\hat{c}_i(t)\,,\\
      \hat{G}_{ij}^>(t,t')&\coloneqq \frac{1}{\mathrm{i}\hbar}\hat{c}_i(t)\hat{c}^\dagger_j(t')\,. \label{eq:G</>_op_def}
\end{align}
Additionally, we will only consider $G^\gtrless$ on the time-diagonal $(t=t')$ and, therefore, denote $G^\gtrless(t)\coloneqq G^\gtrless(t,t)$. On the real-time diagonal, the lesser component of the one-body Green function is proportional to the single-particle density matrix, $n_{ij}(t)\coloneqq \langle \hat{c}^\dagger_j(t)\hat{c}_i(t)\rangle=\pm\mathrm{i}\hbar G^<_{ij}(t)$. In this paper, we will not consider bosons in a condensate and thus no anomalous correlators will appear. However, an extension of our approach to that case is straightforward. \\

The cornerstone of the quantum fluctuation approach, as developed in Ref.~\cite{schroedter_cmp_22}, is the single-particle fluctuation operator,
\begin{equation}
    \delta\hat{G}_{ij}(t)\coloneqq \hat{G}_{ij}^<(t)-G^<_{ij}(t)\equiv \hat{G}^>_{ij}(t)-G^>_{ij}(t)\,,
\label{eq:deltag-def}
\end{equation}
where it was used that,  on the time-diagonal, \\ $\hat{G}^>_{ij}(t)-\hat{G}^<_{ij}(t)=\frac{1}{\mathrm{i}\hbar}\delta_{ij}$, for all $t$ and, obviously, $\langle \delta\hat{G}_{ij}(t)\rangle =0$. Next, we define general two-particle fluctuations and the associated correlation function as
\begin{align}
   \hat L_{ijkl}(t,t') &\coloneqq  \delta\hat{G}_{ik}(t)\delta\hat{G}_{jl}(t')\,,
   \label{eq:hat-l_def}\\
    L_{ijkl}(t,t') &\coloneqq \left\langle \hat L_{ijkl}(t,t')\right\rangle\,, \label{eq:generalized_gamma_def}
    \\
    L_{ijkl}(t) &\coloneqq L_{ijkl}(t,t)\,.
\end{align}

The two-particle correlation function, Eq.~\eqref{eq:generalized_gamma_def}, can be considered  a special case of the exchange-correlation function in standard NEGF theory \cite{stefanucci-book},
\begin{align}
    L_{ijkl}(z_1,z_2,z_3,z_4)\coloneqq \,&G^{(2)}_{ijkl}(z_1,z_2,z_3,z_4)\\&-G_{ik}(z_1,z_3)G_{jl}(z_2,z_4)\,, \label{eq:XC-function_def}
\end{align}
where $G^{(2)}$ is the two-particle Green function defined on the Keldysh contour,
\begin{equation}
    G^{(2)}_{ijkl}(z_1,z_2,z_3,z_4)\coloneqq -\frac{1}{\hbar^2}\left\langle \mathcal{T}_\mathcal{C}\left\{\hat{c}_i(z_1)\hat{c}_j(z_2)\hat{c}^\dagger_l(z_4)\hat{c}^\dagger_k(z_3)\right\}\right\rangle\,. \label{eq:G2_def}
\end{equation}
Depending on the index combinations, the function $L$, Eq.~\eqref{eq:generalized_gamma_def}, is related to various correlation functions. In particular, for $i=k$ and $j=l$, it gives access to density fluctuations and the dynamic structure factor \cite{kremp_2005}, whereas other combinations contain information about the current correlations.

%
\subsection{Quantum dynamics in terms of fluctuations} \label{ss:dynamics}
The equation of motion (EOM) for $G^<$ on the time-diagonal can be given in terms of two-particle fluctuations $L$ \footnote{For two single-particle quantities $A$ and $B$, we define the commutator as $[A,B]_{ij}\coloneqq \sum_k \{A_{ik}B_{kj}-B_{ik}A_{kj}\}$.},
\begin{align}
    \mathrm{i}\hbar\frac{\mathrm{d}}{\mathrm{d}t}G^<_{ij}(t)=\Big[h^\hh,G^<\Big]_{ij}(t)+\Big[I+I^\dagger\Big]_{ij}(t)\,,\label{eq:G<_EOM}
\end{align}
where we introduced an effective single-particle Hartree Hamiltonian,
\begin{equation}
    h^\hh_{ij}(t)\coloneqq h_{ij}(t)\pm\mathrm{i}\hbar\sum_{kl}w_{ikjl}(t)G^<_{lk}(t)\,, \label{eq:hartree_hamiltonian_def}
\end{equation}
and a collision term,
\begin{equation}
    I_{ij}(t)\coloneqq \pm\mathrm{i}\hbar \sum_{klp}w_{iklp}(t)L_{plkj}(t)\,. \label{eq:collision_term_def}
\end{equation}
We can equivalently write 
\begin{equation}
    \Big[I+I^{\dagger}\Big]_{ij}(t)=\Big\langle\left[\delta\hat{U}^\hh,\fluc\right]_{ij}(t)\Big\rangle \,,\label{eq:collision_term_2nd}
\end{equation}
with the operator of an effective single-particle Hartree potential induced by fluctuations,
\begin{align}
    \delta\hat{U}^\hh_{ij}(t)\coloneqq \pm\mathrm{i}\hbar\sum_{kl}w_{ikjl}(t)\fluc_{lk}(t)\,. \label{eq:fluc_hartree_potential_def}
\end{align}
Eq.~\eqref{eq:collision_term_def} can also be equivalently expressed in terms of symmmetric two-particle fluctuations, i.e.,
\begin{align}
I_{ij}(t)=S_{ij}(t)+I^{S}_{ij}(t)\,, \label{eq:collision_term_sym}
\end{align}
with the symmetrized collision term given by 
\begin{equation}
    I^S_{ij}(t)\coloneqq \pm\frac{\mathrm{i}\hbar}{2} \sum_{klp}w_{iklp}(t)\big\{L_{plkj}(t)+L_{lpjk}(t)\big\} \label{eq:collision_term_sym_def}
\end{equation}
and a symmetrization contribution of the form
\begin{equation}
    S_{ij}(t)\coloneqq \frac{1}{2}\sum_{kl}w_{kljk}(t)G^<_{il}(t)\,. \label{eq:symmetrization_term_def}
\end{equation}

The EOM for any quantity that depends on products of $\fluc$, such as $L$, can simply be derived from the EOM for the single-particle fluctuation operator which is given by 
\begin{align}
    \mathrm{i}\hbar\frac{\mathrm{d}}{\mathrm{d}t}\fluc_{ij}(t)=&\Big[h^\hh,\fluc\Big]_{ij}(t)+\Big[ \delta\hat{U}^\hh,G^<\Big]_{ij}(t)
    \label{eq:deltaG_EOM}
    \\&+\Big[ \delta\hat{U}^\hh,\fluc\Big]_{ij}(t)-\Big\langle\Big[\delta\hat{U}^\hh,\fluc\Big]_{ij}(t)\Big\rangle \nonumber\,.
\end{align}
Note that Eq.~\eqref{eq:deltaG_EOM} is nonlinear in $\fluc$, which leads,  in the EOMs for $L$, to terms that are cubic in $\fluc$ and, thus, a coupling to three-particle fluctuations. Therefore, we require approximations decoupling the fluctuations hierarchy. 

\subsection{Quantum Polarization Approximation} \label{ss:quantum_polarization_approximation}
The approximation we want to consider here is the quantum analogue of the classical polarization approximation which is known to be equivalent to the Balescu-Lenard kinetic equation which describes scattering of charged particles on a dynamically screened pair potential  \cite{klimontovich_1982}. Additionally, the polarization approximation is the classical limit of the nonequilibrium $GW$-approximation \cite{bonitz_qkt}. On the level of single-particle fluctuations, the quantum polarization approximation (PA) can be derived by assuming 
\begin{align}
    \delta \hat L_{ijkl}(t)&\approx \delta \hat{L}^{(0)}_{ijkl}(t)\,,\label{eq:polarization_approximation} 
\end{align}
where $\delta\hat{L}$ denotes fluctuations of two-particle fluctuations included in the last line of Eq.~\eqref{eq:deltaG_EOM} and $\delta\hat{L}^{(0)}$ can be interpreted as fluctuations of ideal two-particle fluctuations, $L^{(0)}\coloneqq \pm G^> G^<$, defined as 
\begin{align}
    \delta \hat{L}^{(0)}_{ijkl}(t)&\coloneqq \pm \Big\{G^>_{il}(t)\fluc_{jk}(t)+\fluc_{il}(t)G^<_{jk}(t)\Big\}\,. \label{eq:fluctuations_ideal_def}
\end{align}
Applying the PA \eqref{eq:polarization_approximation}, to Eq.~\eqref{eq:deltaG_EOM} leads to the EOM for $\fluc^\mathrm{PA}$ of the form 
\begin{align}
    \mathrm{i}\hbar\frac{\mathrm{d}}{\mathrm{d}t}\fluc^\mathrm{PA}_{ij}(t)=&\Big[h^\HF,\fluc^\mathrm{PA}\Big]_{ij}(t)+\Big[ \delta\hat{U}^\HF,G^<\Big]_{ij}(t) \,,\,
    \label{eq:deltaG_QPA_EOM}
\end{align}
where we introduced the effective single-particle Hartree-Fock Hamiltonian and the operator of an effective single-particle Hartree-Fock potential induced by fluctuations,
\begin{align}
    h^\HF_{ij}(t)&\coloneqq h_{ij}(t)\pm\mathrm{i}\hbar\sum_{kl}w_{ikjl}^\pm(t)G^<_{lk}(t)\,, \label{eq:hartree-fock_hamiltonian_def}\\
    \delta\hat{U}^\HF_{ij}(t)&\coloneqq \pm\mathrm{i}\hbar\sum_{kl}w^\pm_{ikjl}(t)\fluc_{lk}(t)\,, \label{eq:fluc_hartree-fock_potential_def}
\end{align}
with the (anti-)symmetrized interaction tensor, $w^\pm$, defined as 
\begin{equation}
    w^\pm_{ijkl}(t)\coloneqq w_{ijkl}(t)\pm w_{ijlk}(t)\,. \label{eq:as_interaction_tensor_def}
\end{equation}
Although the EOM for $G^<$, cf. Eq.~\eqref{eq:G<_EOM}, depends only on one-time two-particle fluctuations, we will, however, consider the more general set of EOMs for the two-time two-particle fluctuations as this allows for access to two-time observables that can be calculated from $L(t,t')$. 
Using Eq.~\eqref{eq:deltaG_QPA_EOM}, we find for two-time two-particle fluctuations within the framework of the PA EOMs of the form (dropping the superscript ``$\mathrm{PA}$'')
\begin{align}
    \mathrm{i}\hbar\frac{\partial}{\partial t}L_{ijkl}(t,t')&=\big[h^\HF,L\big]^{(1)}_{ijkl}(t,t')+\pi^{(1)}_{ijkl}(t,t')\,,\; \label{eq:ggamma_1_EOM}\\
    \mathrm{i}\hbar\frac{\partial}{\partial t'}L_{ijkl}(t,t')&=\big[h^\HF,L\big]^{(2)}_{ijkl}(t,t')+\pi^{(2)}_{ijkl}(t,t')\,,\;\: \label{eq:ggamma_2_EOM}
\end{align}
where we introduced Hartree-Fock terms of the form
\begin{align}
    \left[h^\mathrm{HF},L\right]^{(1)}_{ijkl}(t,t')&\coloneqq\sum_{p}\big\{h^{\mathrm{HF}}_{ip}(t)L_{pjkl}(t,t')\\&\hspace{10mm}-h^{\mathrm{HF}}_{pk}(t)L_{ijpl}(t,t')\big\} \,,\label{eq:HF_term_1_def}\\
    \left[h^\mathrm{HF},L\right]^{(2)}_{ijkl}(t,t')&\coloneqq\sum_{p}\big\{h^{\mathrm{HF}}_{jp}(t')L_{ipkl}(t,t')\\&\hspace{10mm}-h^{\mathrm{HF}}_{pl}(t')L_{ijkp}(t,t')\big\}\,, \label{eq:HF_term_2_def}
\end{align}
and the polarization terms,
\begin{align}
    \pi^{(1)}_{ijkl}(t,t')&\coloneqq\pm\mathrm{i}\hbar\sum_{pqr}\big\{w^\pm_{ipqr}(t)L_{rjpl}(t,t')G^<_{qk}(t)\\&\hspace{15mm}-w^\pm_{pqkr}(t)L_{rjql}(t,t')G^<_{ip}(t)\big\} \,,\label{eq:pi_ggamma_1_def}\\
    \pi^{(2)}_{ijkl}(t,t')&\coloneqq\pm\mathrm{i}\hbar\sum_{pqr}\big\{w^\pm_{pjqr}(t')L_{iqkp}(t,t')G^<_{rl}(t') \\&\hspace{15mm}-w^\pm_{pqrl}(t')L_{irkp}(t,t')G^<_{jq}(t')\big\}\,. \label{eq:pi_ggamma_2_def}
\end{align}

However, by applying the PA, we find that exchange symmetries of the exact two-particle exchange-correlation function are broken which are essential for energy conservation and the stability of numerical calculations. This problem can be circumvented by considering,  instead, symmetric two-particle fluctuations, i.e., considering Eq.~\eqref{eq:collision_term_sym} instead of Eq.~\eqref{eq:collision_term_def} in Eq.~\eqref{eq:G<_EOM} and symmetrizing the initial conditions for $L$. As \eqrefs{eq:ggamma_1_EOM}{eq:ggamma_2_EOM} are linear in $L$, this symmetrization is sufficient to restore the necessary symmetries lost by introducing the PA. \\ 
As was shown in Ref.~\cite{schroedter_cmp_22}, the PA is equivalent to the $GW$ approximation with exchange corrections, as given within the G1--G2 scheme, see Refs.~\cite{schluenzen_prl_20,joost_prb_20,joost_prb_22}, for weak coupling. Furthermore, it has been shown in Ref.~\cite{schroedter_cmp_22} that assuming 
\begin{align}
    \delta\hat{L}_{ijkl}(t) \approx
    \pm\fluc_{il}(t)G^<_{jk}(t)\,,\label{eq:GW_fluc_approximation}
\end{align}
leads to an approximation equivalent to the G1--G2--$GW$ approximation, in the weak coupling limit. 

\subsection{Density Response Function and Dynamic Structure Factor} \label{s:density_response_function}
There is a variety of observables that depend on fluctuations, which are generally not on the time-diagonal, but depend on multiple (in general) independent time arguments. Two particularly important observables are the density response function and dynamic structure factor. 
Here, we consider the (retarded) density response function which, for an arbitrary basis, is defined as 
\begin{align}
    \chiR_{ij}(t,t')&\coloneqq \mathrm{i}\hbar\Theta(t-t')\left\langle\left[\fluc_{ii}(t),\fluc_{jj}(t')\right]\right\rangle \,\\
    &=\mathrm{i}\hbar \Theta(t-t')\{L_{ijij}(t,t')-L_{jiji}(t',t)\}\\
    &=-2\hbar\Theta(t-t')\mathrm{Im}\big[L_{ijij}(t,t')\big]\,,\label{eq:density_reponse_function_def}
\end{align}
%
where exchanges symmetries of two-particle fluctuations were used, i.e., $L_{ijkl}(t,t')=[L_{lkji}(t',t)]^*$.
Hence, the dynamics of $\chiR$ in the PA is directly given by \eqrefs{eq:ggamma_1_EOM}{eq:ggamma_2_EOM} for two combinations of indices for which $\fluc$ describes density fluctuations.

%
Considering a representation of the system in position space, i.e., $G_{ij}(t)\rightarrow G(\mathbf{r},\mathbf{r}',t)$, we additionally introduce the center-of-mass and relative time and position, i.e.,
\begin{align}
    \tau&\coloneqq t_1-t_2\,, & T&\coloneqq \frac{t_1+t_2}{2}\,,\\
    \mathbf{r}&\coloneqq \mathbf{r}_1-\mathbf{r}_2\,, & \mathbf{R}&\coloneqq \frac{\mathbf{r}_1+\mathbf{r}_2}{2}\,.
\end{align}
With this, two-particle fluctuations can then be expressed, equivalently, as
\begin{equation}
    L(\mathbf{r}_1,\mathbf{r}_2,t_1,t_2)\rightarrow L(\mathbf{r},\mathbf{R},\tau, T)\,.
\end{equation}
Next, the dynamic structure factor is defined as the Fourier transform of the correlation function of fluctuations of the charge density (intermediate scattering function), i.e., we consider the Fourier transform of two-particle fluctuations with respect to the relative time and position
\begin{align}
    S(\mathbf{q},\omega,\mathbf{R},T)\coloneqq \int \int\limits_{-\infty}^\infty L(\mathbf{r},\mathbf{R},\tau, T) e^{\mathrm{i}(\omega\tau-\mathbf{r\cdot\mathbf{q})}}\mathrm{d}\tau\mathrm{d}^3\mathbf{r} \quad \label{eq:dynamic_structure_factor_definition}
\end{align}
For systems in equilibrium, the dynamic structure factor does not depend on the center-of-mass time, $T$, where the same applies for spatially homogeneous systems with respect to the center-of-mass position, $\mathbf{R}$. Analogously, this applies also to the density response function in the general case. \\

In similar manner one can also define the response function and structure factor of spin density fluctuations which, however, will not be considered in this work.

\section{Stochastic Approach to the quantum fluctuation dynamics} \label{s:smf_theory}

\subsection{General Concept} \label{ss:general_concept}
%
%

The stochastic mean-field theory (SMF) was first introduced by Ayik \cite{ayik_plb_08} and later extended by Lacroix and many others, see e.g., Refs.~\cite{lacroix_prb14,lacroix_epj_14}. Here, quantum mechanical operators are replaced by stochastic quantities, i.e., $\fluc_{ij}(t)\rightarrow \cfluc_{ij}(t)$, and the quantum mechanical expectation value by the standard stochastic expectation value, i.e., $\langle\cdot\rangle\rightarrow \overline{(\cdot)}$. In practice, the stochastic expectation value is approximated by the arithmetic mean, where the superscript ``$\lambda$'' denotes a random realization, each of which is generated for the initial state according to a statistical ensemble and then propagated in time. This allows operator equations, e.g., \eqrefs{eq:deltaG_EOM}{eq:deltaG_QPA_EOM}, to be solved approximately. Effectively, this is done at the single-particle level and only involves simple mean-field dynamics. The solution of the many-body problem therefore essentially reduces to the construction of a probability distribution for the initial state, mean-field dynamics and semi-classical averaging \cite{lacroix_prb14}.\\

Replacing products of non-commuting operators with random variables, however, requires symmetrization of said products. The probability distribution describing the initial state at $t=t_0$ is then constructed in such a way that all symmetrized quantum mechanical moments are equal to the semi-classical moments. For the case of an ideal (uncorrelated) state, the first two quantum mechanical moments are given by
\begin{align}
    \langle\fluc_{ij}(t_0)\rangle &=0\,, \label{eq:1st_moment}\\
    L_{ijkl}(t_0)= \langle \fluc_{ik}(t_0)\fluc_{jl}(t_0)\rangle &=-\frac{1}{\hbar^2}\delta_{il}\delta_{jk}n_j(1\pm n_i)\,, \quad \label{eq:2nd_moment}
\end{align}
with $n_i\coloneqq \pm \mathrm{i}\hbar G^<_{ii}(t_0)$. We underline that considering only an ideal initial state is not a restriction because a correlated initial state can be produced from an ideal state via the adiabatic switching method, e.g. Ref.~\cite{hermanns_prb14,schluenzen_jpcm_19}. It has to be noted, however, that no probability distribution exists which reproduces all symmetrized quantum mechanical moments \cite{Lacroix2019}.


\subsection{Stochastic Polarization Approximation} \label{ss:SPA}
We now apply the SMF approach to the PA, i.e., Eq.~\eqref{eq:deltaG_QPA_EOM}, and find the stochastic polarization approximation (SPA), which is of the form (we drop the superscript ``$\mathrm{SPA}$'' below and imply that all further considerations within the framework of the SMF theory are done using the SPA)
\begin{align}
    \mathrm{i}\hbar\frac{\mathrm{d}}{\mathrm{d}t}\cfluc_{ij}(t)=\Big[h^\HF,\cfluc\Big]_{ij}(t)+\Big[\Delta U^{\HF,\lambda},G^<\Big]_{ij}(t)\,, \label{eq:deltaG_SPA_EOM}
\end{align}
where we introduced the SMF-analogue of the effective single-particle Hartree-Fock potential induced by fluctuations, cf. Eq.~\eqref{eq:fluc_hartree-fock_potential_def},
\begin{equation}
    \Delta U^{\HF,\lambda}_{ij}(t)\coloneqq \pm\mathrm{i}\hbar\sum_{kl}w_{ikjl}^\pm(t)\cfluc_{lk}(t)\,.
\end{equation}
Applying the PA to the SMF theory leads to a neglect of any coupling to higher moments and thereby reduces the impact of choosing an approximate probability distribution for the initial state. In fact, extensive calculations within the framework of the Hubbard model indicate that all distributions are equivalent, provided they correctly describe the first two moments. Analogously, we can derive a stochastic version of the $GW$ approximation (SGW) within the fluctuations framework for weak coupling \cite{schroedter_cmp_22}.\\

The numerical scaling (CPU time and memory) of this approximation is proportional to the number of samples $K$. The CPU-time scaling of the SPA/SGW is given by $\mathcal{O}(KN_b^4N_t)$, where $N_b$ the number of basis states and $N_t$ the number of time steps. Due to this dependence, the specific choice of sampling allows for optimization and can make this approach advantageous compared to the G1-G2-$GW$ approximation which has a scaling of $\mathcal{O}(N_b^6N_t)$ \cite{joost_prb_20}.  

\subsection{Multiple Ensembles Approach} \label{s:ME}
Although the SMF approach allows for a solution of certain operator equations and provides several other advantages compared to other methods, there are significant shortcomings of this theory that have to be addressed. The most striking is the semiclassical nature of the approach, i.e., the attempt to describe a quantum mechanical system using a classical probability distribution. Even if the SPA restricts the number of significant moments, this stochastic approximation poses major limitations as quantum coherence effects cannot be properly accounted for, thus, restricting this approach to weakly and moderately coupled systems. Additionally, this approach is unable to compute any observables that depend on the ordering of the underlying operators, e.g., the density response function, cf. Eq.~\eqref{eq:density_reponse_function_def}.
Here, we introduce a partial solution to this problem within the framework of the SPA (or equivalently any approximation to the fluctuations hierarchy that only takes into account the first two moments).

\subsubsection{Basic Ideas} \label{ss:basic_ideas}
The solution we propose is the multiple ensembles (ME) approach. Instead of considering only one statistical ensemble to describe realizations of the initial state, we consider two equivalent ensembles, and replace quantum fluctuations according to
\begin{align}
    \fluc_{ij}\rightarrow (\Delta G^{(1),\lambda}_{ij},\Delta G^{(2),\lambda}_{ij})\,,\label{eq:ME_assumption_1}
\end{align}
whereas products of operators are replaced based on their ordering:
\begin{align}
\hat L_{ijkl} = \fluc_{ik}\fluc_{jl}\rightarrow \Tilde{L}^\lambda_{ijkl}\coloneqq \Delta G^{(1),\lambda}_{ik}\Delta G^{(2),\lambda}_{jl}\,. \label{eq:ME_assumption_2}
\end{align}
Analogously, we define the expectation value of two-particle fluctuations within the ME approach as 
\begin{align}
    \Tilde{L}_{ijkl}(t,t')&\coloneqq \overline{\Tilde{L}^\lambda_{ijkl}(t,t')}
    \,, \label{eq:two-particle_fluctuations_ME_def} \\
    \Tilde{L}_{ijkl}(t)&\coloneqq \Tilde{L}_{ijkl}(t,t')\,.
\end{align}
Here, we see why this approach is restricted to approximations that only consider the first two moments. For example, the exact EOM for $\fluc$, cf. Eq.~\eqref{eq:deltaG_EOM}, includes terms which are quadratic in $\fluc$. These terms would then be replaced by scalar quantities, cf. Eq.~\eqref{eq:ME_assumption_2}, whereas all terms that are only linear in $\fluc$ would be replaced by two component quantities, cf. Eq.~\eqref{eq:ME_assumption_1}. Additionally, we are unable to properly define third and higher moments within this approach. This is, however, not a problem within the framework of the SPA since higher moments are neglected, and all considered observables only depend on two-particle fluctuations, including the correlation energy and the density response function. \\

Since we do not require any additional symmetrization, by applying the ME, compared to standard SMF theory, the constraints for the initial state are given by replacing the operators in \eqrefs{eq:1st_moment}{eq:2nd_moment} according to \eqrefs{eq:ME_assumption_1}{eq:ME_assumption_2},
\begin{align}
    \overline{\Delta G^{(1),\lambda}_{ij}(t_0)}=\overline{\Delta G^{(2),\lambda}_{ij}(t_0)}&=0\,, \label{eq:1st_moment_ME}\\
    \overline{\Delta G^{(1),\lambda}_{ik}(t_0)\Delta G^{(2),\lambda}_{jl}(t_0)}&=-\frac{1}{\hbar^2}\delta_{il}\delta_{jk}n_j(1\pm n_i)\,.\quad \label{eq:2nd_moment_ME}
\end{align}
The single-particle fluctuation operator, $\fluc$, obeys the symmetry 
\begin{equation}
    \fluc_{ij}=-\fluc_{ji}^\dagger\,. \label{eq:deltaG_sym}
\end{equation}
Within the standard SMF approach, we construct the random variables so that this symmetry is preserved, i.e., $\cfluc_{ij}=(-\cfluc_{ji})^*$. For the ME approach, however, we demand, instead, that the fluctuations obey the symmetry 
\begin{equation}
    \Delta G^{(1),\lambda}_{ij}\equiv -(\Delta G^{(2),\lambda}_{ji})^*\,. \label{eq:deltaG_ME_sym}
\end{equation}
Thus, the newly defined two ensembles reduce to a single random variable, and we define $\cfluc_{ij}\coloneqq \Delta G^{(1),\lambda}_{ij}\equiv -(\Delta G^{(2),\lambda}_{ji})^*$. Imposing this relation we, therefore, find for two-particle fluctuations within the ME approach,
\begin{equation}
    \Tilde{L}_{ijkl}(t,t')=\big[\Tilde{L}_{lkji}(t',t)\big]^*\,, \label{eq:two-particle_fluctuations_symmetry_ME}
\end{equation}
thus reproducing one of the exchange properties of the exact two-particle fluctuations. Furthermore,
\eqrefs{eq:1st_moment_ME}{eq:2nd_moment_ME} take to following form,
\begin{align}
    \overline{\cfluc_{ij}(t_0)}&=0\,,\label{eq:1st_moment_ME_2}\\
    \overline{\cfluc_{ik}(t_0)(\cfluc_{lj}(t_0))^*}&=\frac{1}{\hbar^2}\delta_{il}\delta_{jk}n_j(1\pm n_i)\,.\label{eq:2nd_moment_ME_2}
\end{align}
Again, we only consider the ME approach from this point onward, unless stated otherwise.

\subsection{Semiclassical dynamics of the fluctuations} \label{ss:dynamics_ME}
The dynamics of $G^<$ and $\cfluc$, within the ME approach, directly follow from \eqrefs{eq:G<_EOM}{eq:deltaG_QPA_EOM} with the symmetric collision term [cf. Eq.~\eqref{eq:collision_term_sym}],
\begin{align}
    \mathrm{i}\hbar\frac{\mathrm{d}}{\mathrm{d}t}G^<_{ij}(t)=&\Big[h^\hh, G^<\Big]_{ij}(t)+\Big[S+S^\dagger\Big]_{ij}(t)\\&+ \Big[I^\mathrm{ME}+I^{\mathrm{ME}\dagger}\Big]_{ij}(t)\,, \label{eq:G^<_ME_EOM}\\
    \mathrm{i}\hbar\frac{\mathrm{d}}{\mathrm{d}t}\cfluc_{ij}(t)=&\Big[h^\HF,\cfluc\Big]_{ij}(t)+\Big[\Delta U^\HF,G^<\Big]_{ij}(t)\,, \label{eq:deltaG_ME_EOM}
\end{align}
where we defined the ME-collision term, $I^\mathrm{ME}$, as 
\begin{align}
    I^\mathrm{ME}_{ij}(t)\coloneqq \pm \frac{\mathrm{i}\hbar}{2}\sum_{klp}w_{iklp}(t)&\big\{ \Tilde{L}_{plkj}(t)+\Tilde{L}_{lpjk}(t)\big\}\,. \label{eq:collision_term_ME_def}
\end{align}
Notice that we again have to resort to symmetrized expressions (anticommutator) due to the symmetry breaking of the underlying approximation.

\subsection{Sampling} \label{ss:Sampling_ME}
The sampling methods for the ME approach are analogous to those of the standard SMF approach. Here, we focus on two approaches to sampling the initial state: stochastic and deterministic sampling that are briefly explained in the the following. 

\subsubsection{Stochastic Sampling} \label{sss:stochastic_sampling}
The standard approach for the construction of the initial state, within the standard SMF theory, is stochastic sampling. Here, known probability distributions, such as a Gaussian distribution or a uniform distribution, are chosen to reproduce the symmetric moments. For the ME approach, we consider a set of independent complex random variables $\cfluc_{ij}(t_0)$ with zero mean and variance given by
\begin{align}
   \overline{|\cfluc_{ij}(t_0)|^2}=\frac{1}{\hbar^2}n_j(1\pm n_i) \,.\label{eq:variance_ME}
\end{align}
As mentioned in Sec.~\ref{ss:general_concept}, the expectation value is usually approximated by the arithmetic mean, i.e., a sufficiently large number of random realizations of the initial state is generated according to said constraints, cf. Eq.~\eqref{eq:variance_ME}. 

\subsubsection{Deterministic Sampling} \label{sss:deterministic_sampling}
In Ref.~\cite{schroedter_cmp_22}, a new sampling method was proposed for the special case of fermions at zero temperature \footnote{An extension to systems with $T>0$ and bosonic systems is straightforward.}. The idea of the so-called ``deterministic sampling'' is to consider a system of nonlinear equations and construct a solution that exactly satisfies the properties of the initial state. The algorithm of Ref.~\cite{schroedter_cmp_22} has to be adjusted only slightly for the ME approach. It follows that the system of nonlinear equations is given by 
\begin{align}
    \sum_{\lambda =1}^{M} \Delta n^{\lambda}_{ij}&= 0\,, \label{eq:exact_sampling_1st_ME}\\
    \sum_{\lambda =1}^{M} \Delta n^{\lambda}_{ik} (\Delta n^{\lambda}_{lj})^* &= M\delta_{il}\delta_{jk}\delta_{n_j,1}\delta_{n_i,0}\,, \label{eq:exact_sampling_2nd_ME}
\end{align}
with $\Delta n^{\lambda}_{ij}\coloneqq -\mathrm{i}\hbar\Delta G^{\lambda}_{ij}(t_0)$ and $M\in \mathbb{N}$. The parameter $M$ has to be chosen such that a solution of the system of equations exists. \\

Here, we consider a fermionic system with spin configurations $\uparrow,\downarrow$ \footnote{This approach can be easily generalized to arbitrary spin configurations.} and assume spin symmetry of the initial state, i.e., $n_i^\uparrow=n_i^\downarrow$. Let $N_b\in\mathbb{N}$ denote the size of the basis for one spin component  and $N_p,N_h\in\mathbb{N}$ denote the number of occupied and unoccupied orbitals, respectively, for one spin component, i.e., $N_p+N_h=N_b$. Without loss of generality, we assume $n_i^\uparrow=n_i^\downarrow=1$ for $i=1,\dots, N_p$ and $n_i^\uparrow=n_i^\downarrow=0$ for $i=N_p+1,\dots, N_b$. A solution to \eqrefs{eq:exact_sampling_1st_ME}{eq:exact_sampling_2nd_ME} can be found analogously to Ref.~\cite{schroedter_cmp_22} by setting $K=4M=4N_pN_h$ and
\begin{align}
    \Delta n^{\alpha\beta,\sigma} &= \begin{pmatrix} 0 & \mathcal{A}^{\alpha\beta,\sigma} \\ 0 & 0\end{pmatrix} 
\end{align}
with $\mathcal{A}^{\alpha\beta,\sigma} \in \CC^{N_h\times N_p}$. The matrix $\mathcal{A}^{\alpha\beta,\sigma}$ is of the form
\begin{align}
     \mathcal{A}_{ij}^{\alpha\beta,\uparrow} &\coloneqq \begin{cases}(-1)^\beta (M/2)^{1/2}& \mbox{for } \varphi(i,j)=\alpha,\, \beta\in\{1,2\}\,, \\ 0 & \mbox{else}, \end{cases}\\
        \mathcal{A}_{ij}^{\alpha\beta,\downarrow} &\coloneqq \begin{cases}(-1)^\beta (M/2)^{1/2}& \mbox{for } \varphi(i,j)=\alpha,\, \beta\in\{3,4\}\,, \\ 0 & \mbox{else}, \end{cases}
\end{align}
for $(\alpha,\beta)\in \{1,\dots, M\}\times\{1,\dots,4\}$ and an arbitrary bijection $\varphi:\{1,\dots,N_p\}\times \{1,\dots, N_h\}\rightarrow \{1,\dots,M\}$. Using this approach for the construction of a solution, we only require a total of $8N_pN_h$ samples, thus deterministic sampling for the ME approach has an advantageous numerical scaling compared to the standard deterministic sampling which requires $16N_pN_h$ samples.

\subsection{Application to the Density Response Function and Dynamic Structure Factor}
\label{ss:application_density_response_function_ME}
We  now apply the ME approach to the density response function,  Eq.~\eqref{eq:density_reponse_function_def}, and find 
\begin{align}
    \chi_{ij}^{\mathrm{R,ME}}(t,t')\coloneqq& -\mathrm{i}\hbar\Theta(t-t')\Big\{\overline{\cfluc_{ii}(t)(\cfluc_{jj}(t'))^*}\\&\hspace{18mm}-\overline{(\cfluc_{ii}(t))^*\cfluc_{jj}(t')}\Big\} \\
    =&2\hbar\Theta(t-t')\mathrm{Im}\Big[\overline{\cfluc_{ii}(t)(\cfluc_{jj}(t'))^*}\Big]\\
    =&-2\hbar\Theta(t-t')\mathrm{Im}\Big[\Tilde{L}_{ijij}(t,t')\Big]\,,\label{eq:density_response_function_ME}
\end{align}
thus reproducing the result of Eq.~\eqref{eq:density_reponse_function_def}.\\

Because we require the two ensembles to satisfy Eq.~\eqref{eq:deltaG_ME_sym}, realizations of single-particle fluctuations are not necessarily purely imaginary, thus, implying that the imaginary part of products of fluctuations can be nonzero, in general. If we required each ensemble to be anti-Hermitian, i.e., $\Delta G^{(n),\lambda}_{ij}=-(\Delta G^{(n),\lambda}_{ji})^*$, for $n\in\{1,2\}$, the diagonal elements would be purely imaginary for arbitrary times. This implies that these products would be real, hence, the retarded density response function would be purely imaginary. The exact density response function, however, is real~\cite {stefanucci-book}. \\

Analogously to previous considerations, the dynamic structure factor follows  within the ME approach, cf. Eq.~\eqref{eq:dynamic_structure_factor_definition}. 
This multiple ensembles approach thus provides direct access to spectral two-particle quantities. In equilibrium, the result depends only on the time difference, $\tau$. However, our approach remains fully valid in nonequilibrium, where the density response, in addition, depends on the center-of-mass time, $T$.

\subsection{Application to the Fermi-Hubbard Model} \label{s:application_to_Hubbard}
\subsubsection{Hubbard Hamiltonian} \label{ss:Hubbard_Hamiltonian}
For the Fermi-Hubbard model, the general pair-interaction of Eq.~\eqref{eq:generic_Hamiltonian} transforms into
\begin{equation}
    w_{ijkl}^{\alpha\beta\gamma\delta}=U\delta_{ij}\delta_{ik}\delta_{il}\delta_{\alpha\gamma}\delta_{\beta\delta}(1-\delta_{\alpha\beta})\,,
\end{equation}
with the on-site interaction $U$ and the spin components being denoted by greek indices. Additionally, the kinetic energy is replaced by a hopping Hamiltonian 
\begin{equation}
    h_{ij}=-\delta_{\langle i,j\rangle}J\,,
\end{equation}
which includes nearest-neighbor hopping, i.e., $\delta_{\langle i,j\rangle}=1$, if the sites $i$ and $j$ are adjacent, and $\delta_{\langle i,j\rangle}=0$, if they are not. The total Hamiltonian is then given by
\begin{equation}
    \hat{H}=-J\sum_{\langle i,j\rangle}\sum_{\sigma}\hat{c}^\dagger_{i\sigma}\hat{c}_{j\sigma}+U\sum_i \hat{n}_i^\uparrow\hat{n}_i^\downarrow  +\hat H^x(t)\,,
\end{equation}
where $\hat H^x$ describes a possible external excitation of the system. Here, we only consider one-dimensional chains and choose two types of excitations. The first is a potential ``kick'' applied to the first site:
\begin{align}
    \hat{H}^x_I(t) = H_0(t-t_0) \left(\hat{n}^\uparrow_1+\hat{n}^\downarrow_1\right)\,, 
    \label{eq:h-excitation}
\end{align}
where $H_0(t-t_0)$ describes a narrow pulse at time $t=t_0$ with amplitude $H_0$. 
The second is a confinement ``quench'' applied to the central part of the chain \cite{schluenzen_prb16}:
\begin{align}
    \hat{H}^x_{II}(t) = V_0(t-t_0)\sum_{i=1}^M \left(\hat{n}^\uparrow_i+\hat{n}^\downarrow_i\right)\,, 
    \label{eq:h-quench}
\end{align}
consisting of $M$ connected sites where we will use $M=N/2$. At time $t=t_0$ the potential is switched off which initiates a diffusion-type process.

\subsubsection{Implementation of SPA-ME} \label{ss:implementation_SPA}
The EOMs for the single-particle Green function and fluctuations in SPA-ME, \eqrefs{eq:G^<_ME_EOM}{eq:deltaG_ME_EOM}, take the following form,
\begin{align}
    \mathrm{i}\hbar\frac{\mathrm{d}}{\mathrm{d}t} G^{<,\sigma}_{ij}(t)=&\Big[h^{\sigma},G^{<,\sigma}\Big]_{ij}(t)+\Big[I+I^\dagger\Big]^\sigma_{ij}(t),\label{eq:EOM_SPA_G1_Hubbard}\\
     \mathrm{i}\hbar\frac{\mathrm{d}}{\mathrm{d}t} \Delta G^{\lambda,\sigma}_{ij}(t)=&\Big[h^{\sigma},\Delta G^{\lambda,\sigma}\Big]_{ij}(t)\\&+\Big[\Delta U^{\lambda,\sigma},G^{<,\sigma}\Big]_{ij}(t), \label{eq:EOM_SPA_deltaG_Hubbard}
\end{align}
where the Hartree-(Fock)-Hamiltonian and fluctuation Hartree-Fock-potential in \eqrefs{eq:EOM_SPA_G1_Hubbard}{eq:EOM_SPA_deltaG_Hubbard}  become, in the Hubbard basis (without external excitation), 
\begin{align}
    h_{ij}^{\sigma}(t)&\coloneqq h_{ij}^{\mathrm{HF},\sigma}(t)\equiv  h_{ij}^{\mathrm{H},\sigma}(t)=-\delta_{\langle i,j\rangle}J-\mathrm{i}\hbar\delta_{ij}U G^{{\overline{\sigma}}}_{ii}(t), \label{eq:hartree-fock-hamiltonian_Hubbard} \\
    \Delta U^{\lambda,\sigma}_{ij}(t)&\coloneqq\Delta U_{ij}^{\HF,\lambda,\sigma}(t)\equiv\Delta U_{ij}^{\mathrm{H},\lambda,\sigma}(t)=-\mathrm{i}\hbar\delta_{ij}U\Delta G^{\lambda,{\overline{\sigma}}}_{ii}(t). \label{eq:hartree-fock-potential_SPA_Hubbard}
\end{align}
Here, $\sigma=\uparrow(\downarrow)$ implies $\overline{\sigma}=\downarrow(\uparrow)$. 
This shows that all exchange contributions 
vanish, due to the specific choice of the pair-interaction so that, in this case, the SPA is equivalent to SGW. 
The collision term in Eq.~\eqref{eq:EOM_SPA_G1_Hubbard} takes the form 
\begin{equation}
    I^{\sigma}_{ij}(t)=-\frac{\mathrm{i}\hbar}{2} U \left\{\Tilde{L}^{\overline{\sigma}\sigma}_{iiij}(t)+\Tilde{L}^{\sigma\overline{\sigma}}_{iiji}(t)\right\}\,, \label{eq:collision_term_SPA_Hubbard}
\end{equation}
whereas the contributions due to symmetrization become
\begin{equation}
    S_{ij}^\sigma(t)=\frac{1}{2}UG^{\sigma}_{ij}(t)=- S_{ij}^{\dagger,\sigma}(t).
\end{equation}
Thus, symmetrization also does not lead to any additional contributions in the Hubbard basis.
Including the external excitation, cf. Eq.~\eqref{eq:h-excitation}, effectively leads to a modification of the on-site interaction of the form $U\rightarrow U+H^x_{i}(t)$, i.e., the on-site interaction $U$ increases for the first site at $t=t_0$ by the amplitude $H_0$.


The initial state of the system in the natural orbital basis of $n(t_0)$ is chosen such that 
\begin{align}
    G_{ij}^\sigma(t_0) &= - \frac{1}{\mathrm{i}\hbar}\delta_{ij}n^\sigma_{i},\\
    \overline{\Delta G_{ij}^{\lambda,\sigma}(t_0)} &= 0,\\
    \overline{\Delta G^{\lambda,\sigma}_{ij}(t_0)(\Delta  G^{\lambda,\sigma'}_{lk}(t_0))^*}&=\frac{1}{\hbar^2}\delta_{il}\delta_{jk}\delta_{\sigma\sigma'}\delta_{n^\sigma_j,1}\delta_{n^{\sigma}_i,0}\,.
\end{align}
Depending on the system's configuration, it becomes necessary to perform a transformation from the natural orbital basis to the Hubbard basis. This can be achieved by diagonalization of the Hamiltonian and the transformation of $G$ and $\Delta G^\lambda$ using its eigenvectors. 

In general, it is necessary to compute a nontrivial interacting ground state from which the externally driven dynamics start. Here, this is done using the so-called ``adiabatic switching method'' \cite{schluenzen_prb16} by replacing the on-site interaction $U$ with a time-dependent interaction $U(t)$. Calculations then start at $t_s$ with an uncorrelated ground state, with $U(t_s)=0$. The on-site interaction is then increased monotonically and sufficiently slowly such that, at $t_0$, the system is in a fully correlated ground state with $U(t_0)=U$. \\

Our main focus will be on the investigation of the density response function, $\chiR$ [cf. Eq.~\eqref{eq:density_response_function_ME}], which is, within the framework of the ME approach, given by
\begin{align}
    \chi^{\mathrm{R,ME}}_{ij}(t,t')= -2&\hbar \Theta(t-t')
    \sum_{\sigma\sigma'}\mathrm{Im}\big[\Tilde{L}_{ijij}^{\sigma\sigma'}(t,t')\big]\,.\label{eq:density_reponse_function_def_hubbard}
\end{align}
By again considering the relative time $\tau$ and center-of-mass time $T$, the Fourier transform of the density response function with respect to $\tau$ is given by 
\begin{equation}
    \chi^{\mathrm{R,ME}}_{ij}(\omega, T)=\int_{-\infty}^\infty \chi^\mathrm{R,ME}_{ij}(\tau, T) e^{\mathrm{i}\omega\tau}\mathrm{d}\tau \,. \label{eq:density_reponse_function_hubbard_FT}
\end{equation}
The spatial coordinate of a lattice site can be defined as $x_i\coloneqq a_0i$ for a one dimensional chain with $a_0$ denoting the characteristic distance of two adjacent sites \footnote{We assume equidistant sites throughout our further considerations.}. Additionally, we now consider equilibrium and periodic boundary conditions (PBC) so that there is no center-of-mass dependence for time and space. We then also consider a Fourier transform with respect to the spatial relative coordinate, i.e., 
\begin{equation}
    \chi^{\mathrm{R,ME}}(q,\omega) =\int \chi^\mathrm{R,ME}_{ij}(\omega)e^{-iqr_{ij}}\mathrm{d}r_{ij}\,,
\end{equation}
with $r_{ij}\coloneqq x_{i}-x_j$. Analogously, we calculate the dynamic structure factor for said system within the ME approach as
\begin{align}
    S^\mathrm{ME}(q,\omega)&=\sum_{\sigma\sigma'} \int\int_{-\infty}^\infty \Tilde{L}_{ijij}^{\sigma\sigma'}(\tau)e^{\mathrm{i}(\omega\tau-r_{ij}q)}\mathrm{d}\tau\mathrm{d}r_{ij}\,, \\
    &= - 4\mathrm{Im}\Big[\chi^\mathrm{R,ME}(q,\omega)\Big] \,.\label{eq:chiR_S_relation}
\end{align}
A generalization to nonequiblirium and spatially inhomogeneous systems is straightforward but will not be considered here. 

For the density response of a system in the ground state (or thermodynamic equilibrium) and the dynamic structure factor we set $t'=t_0$ and propagate only $\Delta G^{\lambda,\sigma}_{ij}(t)$. The final step is the averaging over the realizations $\lambda$ to calculate the two-particle fluctuations and all further observables.\\

A different approach to the calculation of the Fourier transform of the density response function as well as the dynamic structure factor is provided by considering a kick excitation of the system, e.g.
to the first site, cf. Eq.~\eqref{eq:h-excitation}, following the idea of Ref.~\onlinecite{kwong_prl_00} as this produces a spectrally broad excitation that excites all transitions that are quantum-mechanically allowed \cite{kwong_prl_00, balzer_epl_12}. In linear response we then find
\begin{align}
n_i(t) &=\int_{t_0}^t  \chi^R_{i1}(t-\bar t)H^x_0 (\bar t) \mathrm{d}\bar t\,, \label{eq:density_linear_response}\\
\chi_{11}(\omega) &=\frac{\tilde n_1(\omega)}{\tilde H_0(\omega)}\,, \label{eq:chi_kick}
\end{align}
where $\Tilde{n}_i(\omega)$ and $\Tilde{H}_0(\omega)$ denote the Fourier transforms of the density on site $i$ and the kick excitation, respectively. The dynamic structure factor then also immediately follows, cf. Eq.~\eqref{eq:chiR_S_relation}.\\

As we are primarily interested in the Fourier transform of the density response function, we introduce a small exponential damping in the results of the time propagation with a factor of $e^{-\eta t}$ to mitigate the influence of the finite propagation length. Although this leads to a broadening of the spectral quantities, this allows for better comparability of the results. The damping constant is chosen such that $e^{-\eta t_\mathrm{max}}<10^{-4}$, where $t_\mathrm{max}$ denotes the maximum propagation time.

    \begin{figure}[h]
    \centering
    \includegraphics[width=\columnwidth]{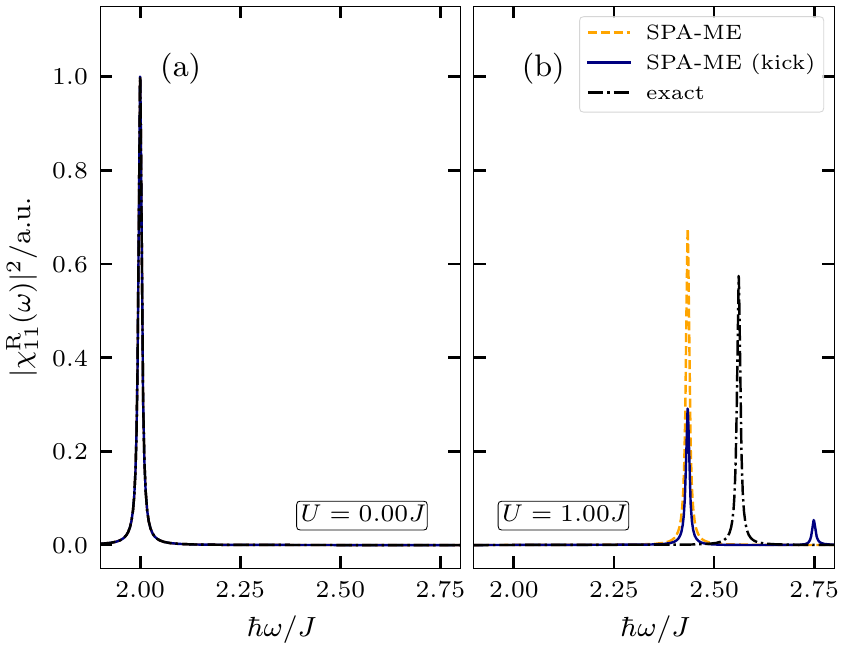}
    \caption{Fourier transform of the density response function (arbitrary units) for the first site of a half-filled Hubbard dimer without PBC, using SPA-ME with (blue lines) and without (orange lines) a kick excitation at $U=0.0J$ (a) and $U=1.0J$ (b). The results are compared to the Fourier transform of the analytical result. The damping constant was chosen to be $\eta =0.005 J/\hbar$.
    }
    \label{fig:spectrum_Ns2}
    \end{figure}
    
    \begin{figure}
    \centering
    \includegraphics[width=\columnwidth]{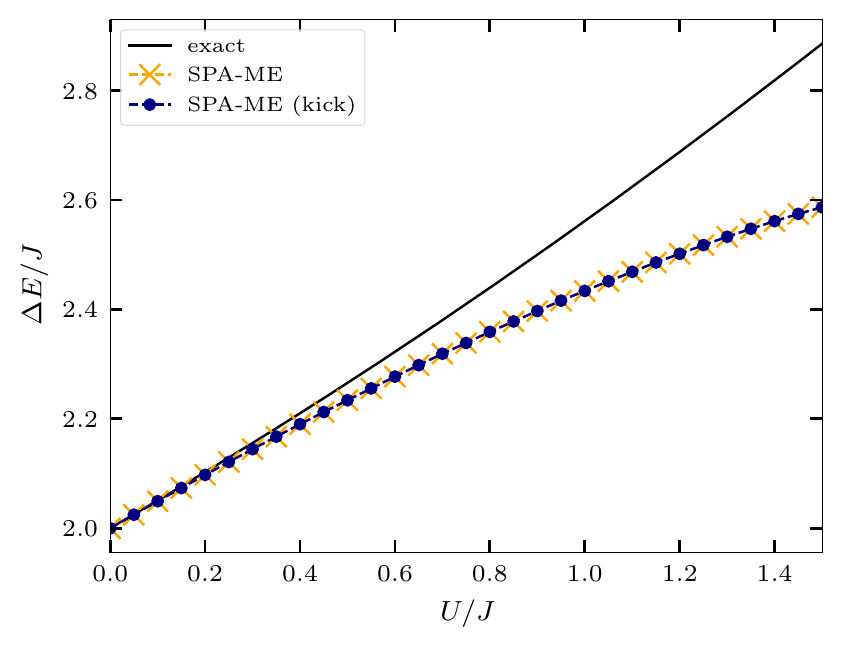}
    \caption{Coupling dependence of the excitation energy of a Hubbard dimer, for the dipole-allowed transition from the ground state, cf. Fig.~\ref{fig:spectrum_Ns2}. Comparison of the present SPA-ME simulations with (blue line) and without (orange line) a kick excitation to the analytical solution. 
    }
    \label{fig:excitation_energy}
    \end{figure}

\section{Numerical results}\label{s:numerics}
We now present numerical results within the stochastic polarization approximation,
systematically extending the results of  Ref.~\cite{schroedter_cmp_22} to response functions. 
First, we consider the ground state and excitation properties of small Hubbard clusters with and without PBC, including the Hubbard dimer and a six-site system, where benchmarks against exact results are possible. We then, in Sec.~\ref{ss:ns=50} turn to larger clusters containing $50$ sites with PBC and compare results for the density response function and dynamic structure factor for SPA-ME and RPA. Finally, in Sec.~\ref{ss:ME_density-response} we study the nonequilibrium dynamics and density response of Hubbard clusters following a short external excitation and a confinement quench.

\subsection{Test of the ME approach for small Hubbard clusters in the ground state} \label{ss:ME-test}
    \begin{figure}[h]
    \centering
    \includegraphics[width=\columnwidth]{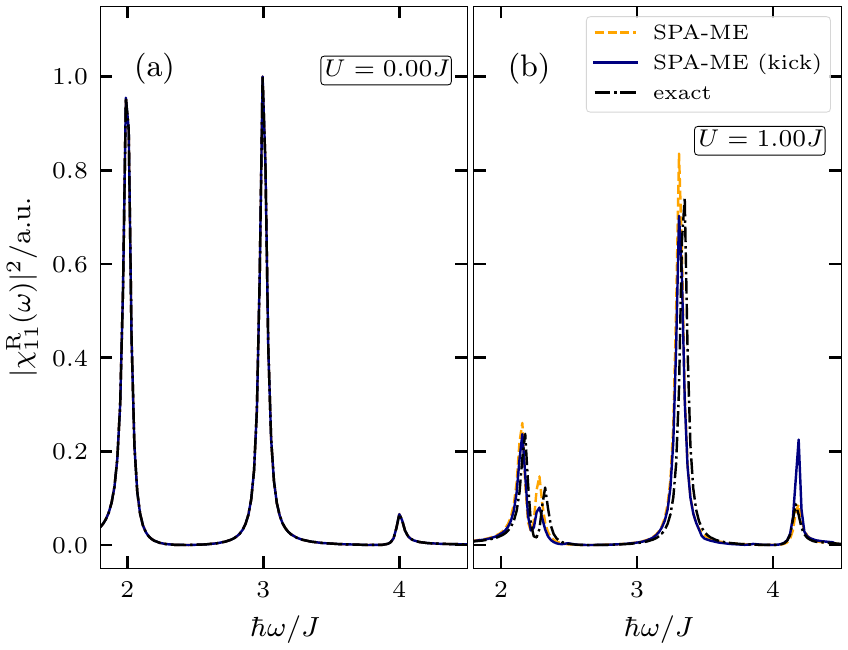}
    \caption{Fourier transform of the density response function (arbitrary units) for the first site of a half-filled six-site system with PBC using SPA-ME with (blue lines) and without (orange lines) kick excitation at $U=0.0J$ (a) and $U=1.0J$ (b). The results are compared to the Fourier transform of an exact diagonalization (CI) calculation for the time-dependent density response function. A damping constant of $\eta= 0.03 J/\hbar$ for the exponential damping of the time propagation was used.
    }
    \label{fig:spectrum_Ns6}
\end{figure}
First, we consider a Hubbard dimer at half-filling without PBC. The peak positions of the Fourier transform of the retarded density response function correspond to energies of particle-hole excitations of the ground state. For the Hubbard dimer, there is only one  dipole-allowed excitation from the ground state with an excitation energy of the form
\begin{equation}
    \Delta E= \frac{U}{2}+\frac{\sqrt{U^2+16J^2}}{2}.\label{eq:delta-e}
\end{equation}
In Fig.~\ref{fig:spectrum_Ns2} the Fourier transform of the ground state result for the retarded density response function at site 1, $\chi_{11}(\omega)$, Eq.~\eqref{eq:density_reponse_function_hubbard_FT}, is shown. Comparison with analytical results confirms that this quantity yields the excitation spectrum of the system which contains a single excitation with the energy $\Delta E$, Eq.~\eqref{eq:delta-e}, which is exactly reproduced in the non-interacting case. Even for $U=1.0J$ the result is in good agreement with the analytical benchmark with regards to the peak position. However, for the relative peak height, significant deviations are visible. As the squared modulus of $\chiR(\omega)$ is proportional to the excitation amplitude, these results imply that SPA-ME overestimates the dipole-allowed transition probability. \\

In Fig.~\ref{fig:excitation_energy} the peak position of the density response function is analyzed in more detail over a broader range of coupling parameters. There are only minor deviations from the exact result, for $U\lesssim 0.5J$. Not surprisingly, these deviations increase monotonically with increasing $U/J$ since SPA is a weak coupling approximation. Also, the functional form of the $U$-dependence deviates from the  analytical solution which increases faster than linear, whereas the slope of the SPA-ME result is sublinear.

\

Next, we extend the analysis to a larger system where the excitation spectrum contains more than one transition. We 
consider a half-filled six-site chain with PBC for the case of a non-interacting system ($U=0.0J$) as well as for moderate coupling ($U=1.0J$). For the non-interacting case we again find excellent agreement between SPA-ME and the exact results which were obtained from CI calculations. Moreover, differences are visible for moderate coupling. We see in Fig.~\ref{fig:spectrum_Ns6} that the peak positions of the present SPA-ME calculations are in very good agreement with exact diagonalization results, however, the peak height of the third peak shows significant deviations, which are less pronounced than for the dimer. The other peaks only display minor deviations and show good qualitative agreement. 
In general, we observe that the quality of our density response results improves with increasing system size \cite{schroedter_cmp_22}.

\subsection{Ground state density response results for large Hubbard clusters} \label{ss:ns=50}
We now turn to larger Hubbard clusters. As an example, we consider a chain with periodic boundary conditions (i.e., a ring), which is half-filled with 50-sites for different coupling strengths.
    \begin{figure}[h]
    \centering
    \includegraphics[width=1.00\columnwidth]{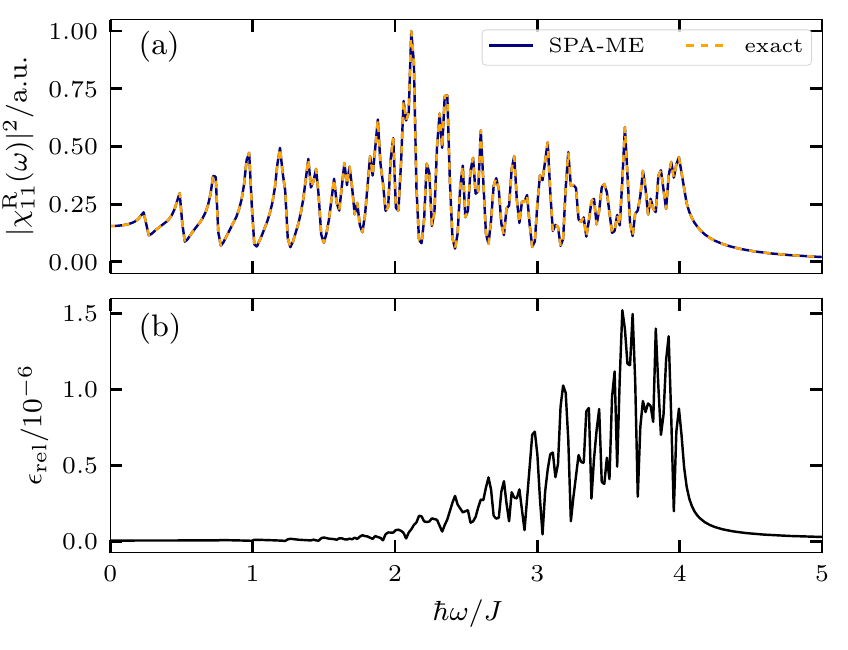}
    \caption{Panel (a): Fourier transform of the density response function (arbitrary units) for the first site of a half-filled $50$-site ring at $U=0.0J$ using SPA-ME. The results are compared to the analytical result. Panel (b): relative error of the Fourier transformed density response.
    A damping constant of $\eta = 0.02 J/\hbar$ was used.
}
    \label{fig:spectrum_Ns50_j0}
\end{figure}
\begin{figure}[h]
    \centering
    \includegraphics[width=\columnwidth]{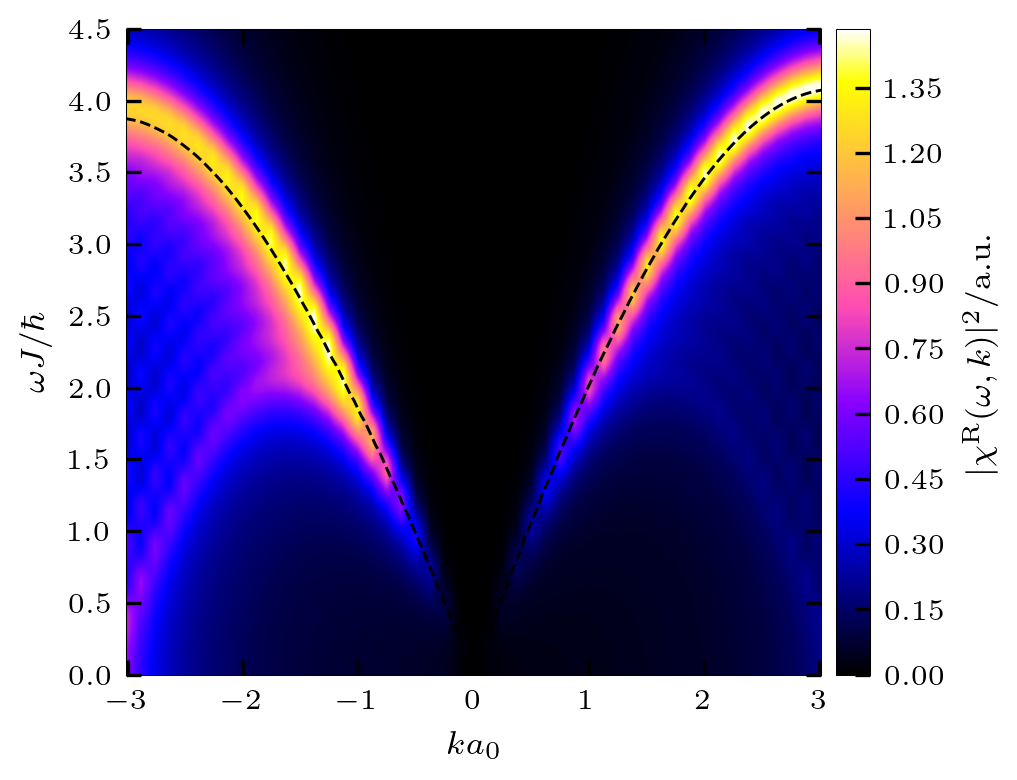}
    \caption{Fourier transform of the density response function (arbitrary units) of a half-filled $50$-site ring  at $U=0.0J$ (left) and $U=1.0J$ (right) using SPA-ME. The dashed lines denote the corresponding peak positions from RPA calculations. A damping constant of $\eta = 0.2 J/\hbar$ was used. 
}
    \label{fig:drf_NS50_Us}
\end{figure}
A similar simulation as for six sites, cf. Fig.~\ref{fig:spectrum_Ns6}, yields for the case of a 50-site ring a much more complex spectrum which is shown in Fig.~\ref{fig:spectrum_Ns50_j0}. We start with the case of a non-interacting system ($U=0.0J$) at half filling and observe excellent agreement with the analytical result for the tight-binding limit as the relative error is of the order of $10^{-6}$.

Next, we consider in Fig.~\ref{fig:drf_NS50_Us} the Fourier transform with respect to the spatial and time coordinate of the density response function from SPA-ME calculations. On the left side the non-interacting system is displayed, while the right side shows the results for a moderately coupled system at $U=1.0J$. Here, we see that the results for the non-interacting case for a finite system with PBC closely resemble the known results for the infinite chain \cite{essler_2005}. The artificial damping of the time propagation leads to a broadening of the peaks due to the finite size of the system, thus causing them to merge into a continuous spectrum. At $U=1.0J$ we see that the previously visible peaks start to vanish whereas the main peaks are slightly shifted upwards. Additionally, Fig.~\ref{fig:drf_NS50_Us} displays the lines of the main peaks for the density response function obtained from RPA calculations. Here, we see that there is excellent agreement between the two approximations with regards to the chosen representation. This comparison of SPA-ME and RPA is further extended in the following.  \\

We now consider the dynamic structure factor of the same system but for varying interaction strengths and specific wave numbers and closely compare results from SPA-ME to RPA calculations.  As the $GW$ approximation for a system in equilibrium is equivalent to the RPA, this is directly extending our comparison of the two approximations from Ref.~\cite{schroedter_cmp_22}. The time-dependent SPA-ME results are multiplied by a factor of $e^{-\eta t}$ with damping constant $\eta = 0.2 J/\hbar$ and the same damping constant, $\eta$, is used for the RPA calculations, see \eqrefs{eq:RPA_density_density_responsefunction}{eq:RPA_polarization_function}. This is done to broaden the $\delta$-like peaks due to the finite system size and better comparability to the infinite chain.

\begin{figure}
    \centering
        \includegraphics[width=\columnwidth]{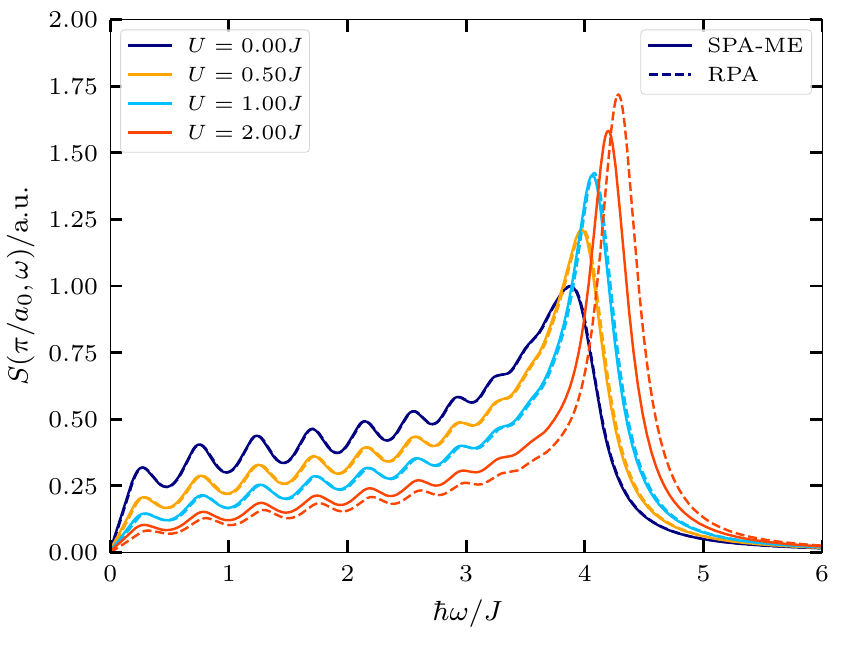}
        \caption{Dynamic structure factor of a half-filled $50$-site ring at \(q=\pi/a_0\) for different \(U\) using SPA-ME (solid lines) and RPA (dashed lines).}
        \label{fig:S_U}
    \end{figure}

First, we consider the dynamic structure factor for a fixed wave number $q=\pi/a_0$ for different interactions strength. Figure~\ref{fig:S_U} shows that there is excellent agreement between the SPA-ME and RPA results, for weak coupling. With increasing interaction strength  deviations become more pronounced, however, only for $U=2.0J$ significant differences between the two approximations are visible. As both approximations are only applicable within the weak coupling regime, this coupling strength is beyond their validity anyway. Moreover, both approximations show a main peak located at $\omega \sim 3.9 J/\hbar$ for $U=0.0 J$ with $S\sim 1.0 \mathrm{a.u.}$. With increasing interaction strength this peak is shifted towards higher frequencies ($\omega \sim 4.2 J/\hbar$) and height with $S\sim 1.6 \mathrm{a.u.}$ at $U=2.0J$, where the RPA peak is shifted to slightly higher frequencies compared to the SPA-ME peak [$S \sim 1.7 \mathrm{a.u.}$ at $\omega \sim 4.3 J/\hbar $]. Additionally, we observe that, due to the finite system size, there are a number of peaks visible next to the main peak which do not exist in the infinite chain. These peaks become less pronounced with increasing interaction strength. Here, the RPA results deviate, for $U=2.0J$, from the SPA-ME results in that there is a slightly larger downward shift visible for the peak height. 

\begin{figure}
    \centering
        \includegraphics[width=\columnwidth]{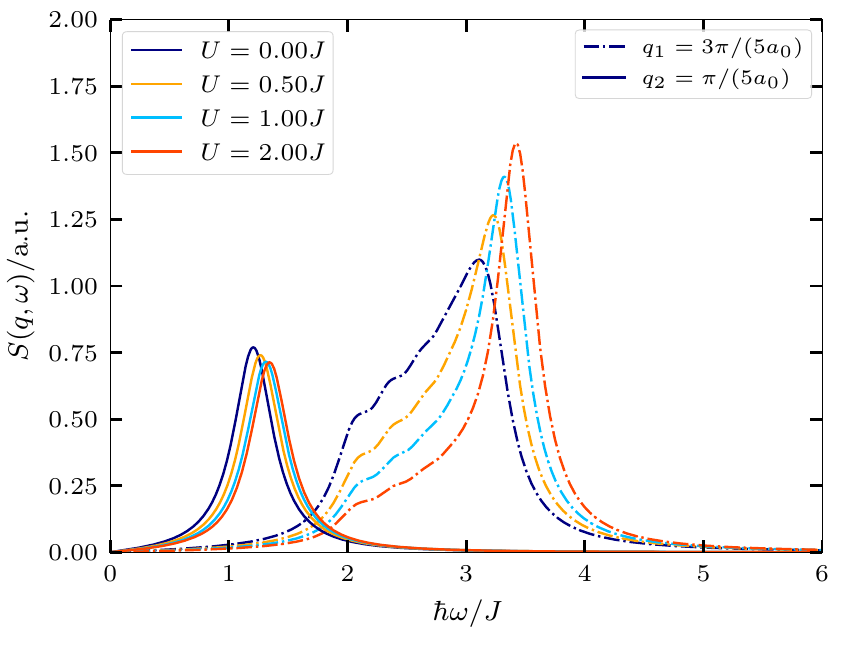}
        \caption{Dynamic structure factor of a half-filled $50$-site ring at \(q_1=3\pi/(5a_0)\) (solid lines) and \(q_2=\pi/(5a_0) \) (dash-dotted lines) for different \(U\) using SPA-ME.}
        \label{fig:S_q0.2/0.6}
    \end{figure}

Next, we consider in Fig.~\ref{fig:S_q0.2/0.6} the dynamic structure factor for two different wave numbers, $q_1= 3\pi/(5a_0)$ and $q_2=\pi/(5a_0)$, at different $U$, using only SPA-ME. For $q_2$ we see that only a single peak is visible for each $U$ located at $\omega \sim 1.2 J/\hbar$ with a height of $S\sim 0.77 \mathrm{a.u.} $ while none of the other peaks due to the finite system size are present. With increasing interaction strength a slight downward shift of the peak height is observed, in contrast to the previous results. Again, however, the peak position is shifted towards higher frequencies. Here, both shifts are not as pronounced as for the previous case. This changes for $q_1$ where the position of the main peak is shifted from $\omega \sim 3.1 J/\hbar$ to $\omega \sim 3.4 J/\hbar$, thus displaying a stronger shift compared to the case $q=\pi/a_0$. Additionally, here the build up of the finite size peaks is observed. However, the peak height is shifted significantly downward, compared to $q_1$ and even $q=\pi/a_0$. Moreover, the peaks become less pronounced with increasing interactions strength so that they are barely visible for $U=2.0J$. Last, the height of the main peak at $U=0.0J$ is located at $S\sim 1.1 \mathrm{a.u.}$ and shifted upwards to $S\sim 1.5\mathrm{a.u.}$ at $U=2.0J$.

\subsection{Nonequilibrium density response following an external kick excitation of a single site} \label{ss:ME_density-response}
After computing the density response for a system in the ground state, we now turn to the time-dependent density response following a rapid weak external perturbation (``kick''). Here, we again consider two approaches: first, computation of the density response function from a ground state calculation with subsequent calculation of the density response in linear response theory according to Eq.~\eqref{eq:density_linear_response}. The second approach is a direct nonequilibrium calculation following the external kick perturbation of the first site, cf. Eq.~\eqref{eq:h-excitation}.

Let us first return to Fig.~\ref{fig:spectrum_Ns2} where the present kick results have been included. Here, we see that, for the non-interacting case, the results from the nonequilibrium SPA-ME calculation agree with both the analytical and the equilibrium SPA-ME results. However, for the moderately coupled system we observe significant deviations, as an additional peak at $\omega= 2.75 J/\hbar$ appears. This effect is known from NEGF calculations \cite{balzer_epl_12}. Moreover, we see that the peak position of the equilibrium approach is well reproduced by one of the peaks from the nonequilibrium calculation. This trend is also visible when considering other interactions strengths, as shown in Fig.~\ref{fig:excitation_energy}. Here, we see only minor deviations between the two approaches. Nonetheless, the height of the main peak shown in Fig.~\ref{fig:spectrum_Ns2} is significantly reduced for the kick results.  \\

Similar behavior is observed for the six-site chain shown in Fig.~\ref{fig:spectrum_Ns6}. Here, we also see that the peak positions are well reproduced by the nonequilibrium simulations. However, significant deviations of the heights of most peaks are visible,  suggesting that the nonequilibrium scenario modifies the transition probabilities of almost all allowed transitions. We observe that, in contrast, the equilibrium results are significantly closer to the amplitudes of the exact solution.\\
\begin{figure}
\centering
\includegraphics[width=\columnwidth]{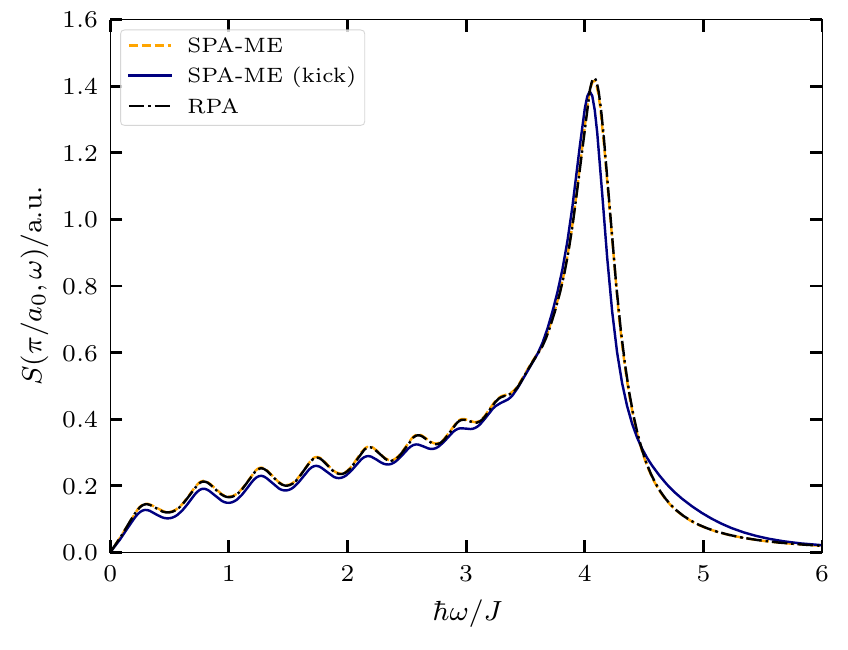}
\caption{Dynamic structure factor for a half-filled $50$-site ring at $U=1.0J$ at \(q=\pi/a_0\) from SPA-ME with (blue line) and without (orange line) a kick excitation and RPA. The time propagated results were exponentially damped with damping constant $\eta =0.2 J/\hbar$.}
\label{fig:S_Ns50_U1.0}
\end{figure}

In Fig.~\ref{fig:S_Ns50_U1.0} we again consider the dynamic structure factor of a 50-site ring. At $U=0.5J$ for a wave number $q=\pi/a_0$, we see the aforementioned excellent agreement of the equilibrium SPA-ME and RPA results. The nonequilibrium SPA-ME calculations display minor deviations compared to the other approaches. Here, the former are shifted slightly downward, for frequencies $\omega \lesssim 3.0J/\hbar$. Additionally, we see that the position of the main peak is shifted towards a smaller frequency with reduced peak height. For larger frequencies ($\omega \gtrsim 4.7J/\hbar$) we observe an increase of the nonequilibrium results compared to the other approaches, which is more pronounced compared to aforementioned shift for smaller frequencies. A possible explanation of these deviations is that nonequilibrium simulations are known to capture correlation effects \cite{kwong_prl_00}, but whether this applies to the present excitation requires further investigation, see also Sec.~\ref{s:discussion}.

\begin{figure}[h]
\centering
\includegraphics[width=\columnwidth]{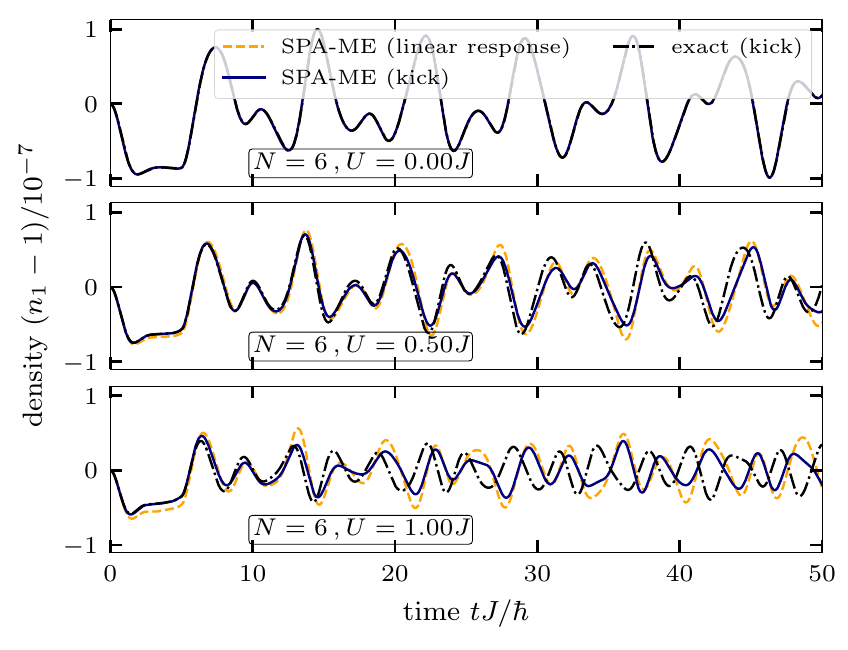}
\caption{Time-dependent density perturbation at the first site for a half-filled six-site chain for three coupling strengths [$U=0.0J$ (a), $U=0.5J$ (b) and $U=1.0J$ (c)] following a kick excitation using SPA (blue lines) and CI. The data are compared to linear response results from a ground state SPA-ME calculation (orange lines), cf. Eq.~\eqref{eq:density_linear_response}. 
}
\label{fig:density_kick}
\end{figure}

Next, we again consider a half-filled six-site chain for different interaction strengths $(U/J=0.0, 0.5, 1.0$) and compare the time-dependent density perturbation at the first site for SPA-ME, from linear response theory and the kick excitation, to CI results. Fig.~\ref{fig:density_kick}  (a) shows that in the non-interacting case both SPA-ME results from linear response and nonequilibrium perfectly agree with the CI data. In (b) we still see good qualitative agreement of both calculations with the CI data with deviations starting to increase for $t\gtrsim 20 \hbar/J$. The linear response results overestimate the amplitude of the oscillations compared to the nonequilibrium calculation. For the moderately coupled system with $U=1.0J$ we immediately see  in (c) that deviations between the results arise. The nonequilibrium SPA-ME results display better agreement with the exact result for longer times ($t\sim 7 \hbar/J$) compared to the linear response results ($t\sim 1 \hbar/J$). Both, however, fail to accurately reproduce the oscillating behavior of the exact result for times larger than $\sim 20 \hbar/J$ as there appears to be a time-dependent phase shift. \\
\begin{figure}[h]
    \centering
    \includegraphics[width=\columnwidth]{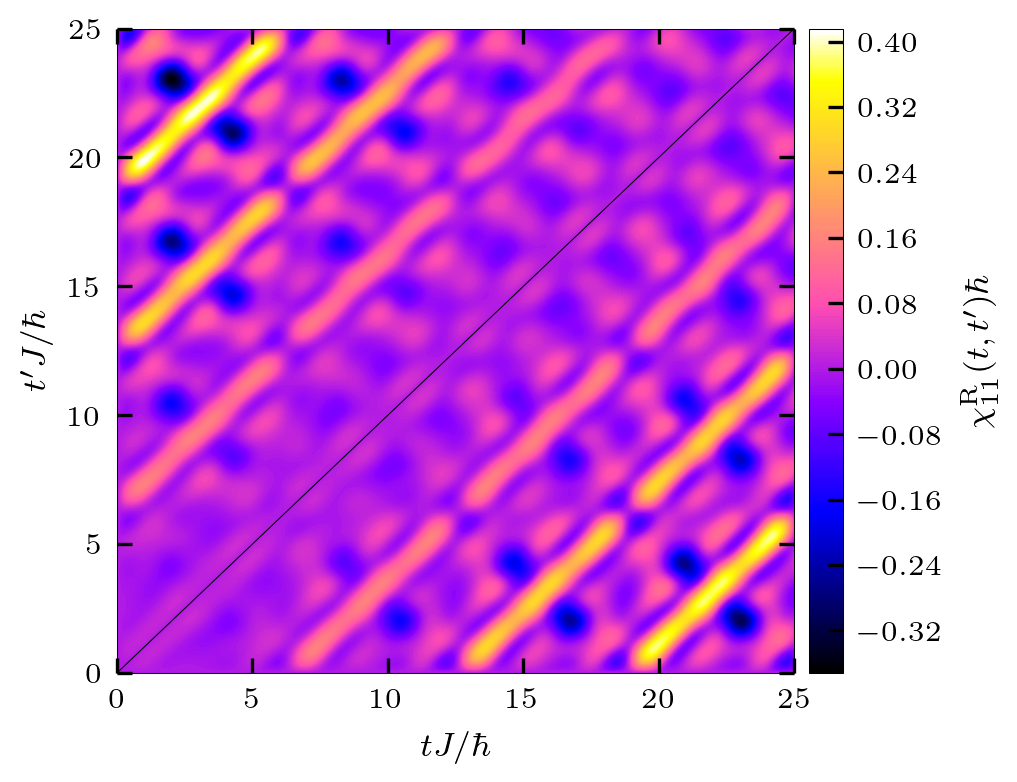}
    \caption{Density response function for a half-filled ring with 6 sites at $U=0.1J$ following a confinement quench at $t=0$. Upper left (lower right) triangle corresponds to SPA-ME (CI) results.}
    \label{fig:neq_drf_ns6_u0.1}
\end{figure}

\subsection{Nonequilibrium density response following a confinement quench} \label{ss:ME_density-response-quench}

As a final nonequilibrium setup, we consider a confinement quench \cite{schluenzen_prb16,schluenzen_prb17} of a half-filled system, cf. Eq.~\eqref{eq:h-quench}. This means, the initial state is such that the left half of the sites is fully occupied, while the right half is empty. Subsequently, at time $T=0$, the confinement potential is removed suddenly, resulting in a rapid expansion of particles in the chain similar to a classical diffusion setup. Thus, immediately after the quench, the system is in strong nonequilibrium, providing an interesting test-bed for our ME approach. In particular, we are interested at this point in the density response during the relaxation process. Again, we start with a small system containing 6 sites with periodic boundary conditions and compare results of SPA-ME with those of exact diagonalization. In Fig.~\ref{fig:neq_drf_ns6_u0.1}, we present the two-time nonequilibrium density response function for the half-filled ring at an interaction strength of $U=0.1J$ obtained from the present SPA-ME model. Recall that the Heaviside function in the definition of the retarded component leads to $\chi^\mathrm{R}_{ij}(t,t')\equiv 0$ for $t'>t$, cf. Eq.~\eqref{eq:density_reponse_function_def}. This allows us to plot in the half plane below the diagonal (black line) the corresponding CI result for the density response function $\chi^\mathrm{R}_{ij}(t',t)$, for the transposed arguments.  Obviously, there is very good agreement between the two results. 
\begin{figure}
    \centering
    \includegraphics[width=\columnwidth]{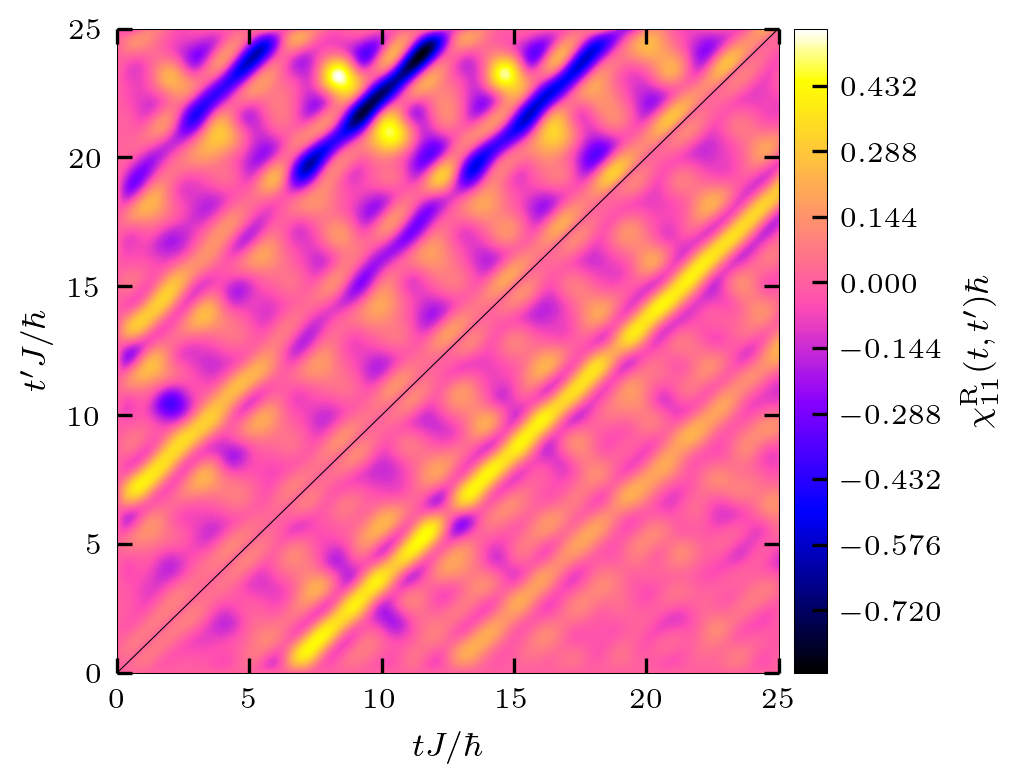}
    \caption{Same as Fig.~\ref{fig:neq_drf_ns6_u0.1}, but at $U=0.4J$}
    \label{fig:neq_drf_ns6_u0.4}
\end{figure}
We now move on to stronger coupling, $U=0.4J$. The results are presented in  Fig.~\ref{fig:neq_drf_ns6_u0.4} where significant differences between the two results are visible. It is noticeable here that the same trends are present for both results with respect to the phase of the oscillations. However, deviations in their amplitudes are visible. Moreover, we observe that the oscillations for the exact result are damped with increasing relative time, while the SPA-ME results illustrate the opposite trend. Comparing the results for the two interaction strengths, we notice that they differ mainly in the amplitude of the oscillations.
\begin{figure*}[t]
    \centering
    \includegraphics[width=2\columnwidth]{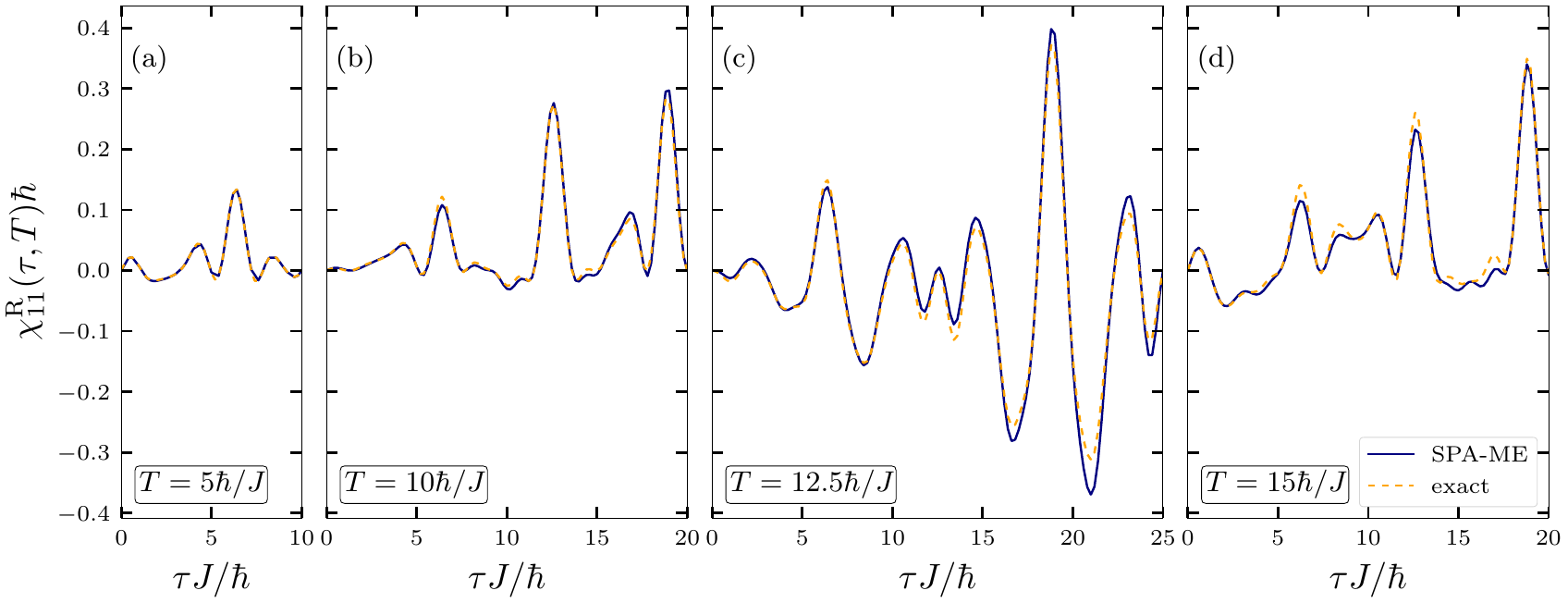}
    \caption{Density response function for different times $T$ following a confinement quench, for a half-filled ring with 6 sites  at $U=0.1J$. Comparison of the present SPA-ME approach to CI calculations.} 
    \label{fig:neq_drf_ns6_u0.1_tau}
\end{figure*}
\begin{figure*}[t]
    \centering
    \includegraphics[width=2\columnwidth]{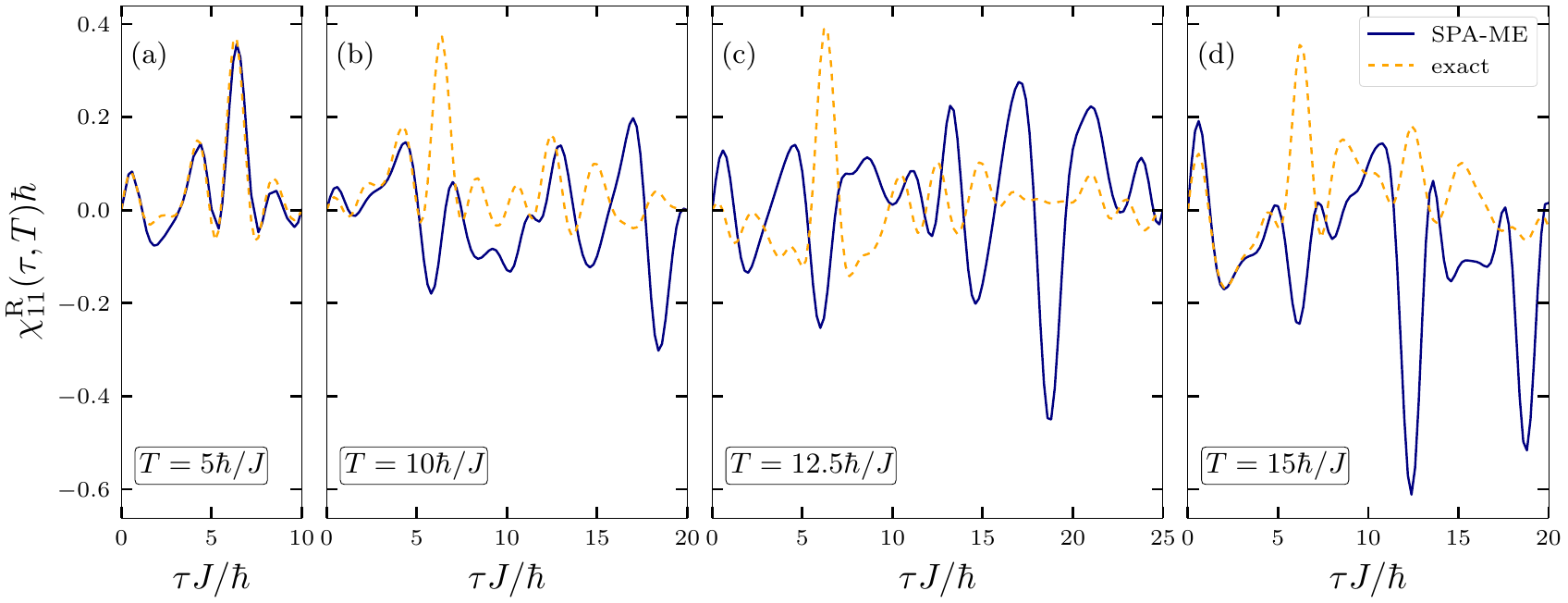}
    \caption{Same as Fig.~\ref{fig:neq_drf_ns6_u0.1_tau}, but at $U=0.4J$.}
    \label{fig:neq_drf_ns6_u0.4_tau}
\end{figure*}

For a better comparison of the CI and SPA-ME data, we now present the results in 2D plots. There we show the density response function vs. relative time, $\tau$, for fixed values of the center-of-mass time, $T$. In Fig.~\ref{fig:neq_drf_ns6_u0.1_tau}  we consider the density response function for four values of $T$ for the case of weak coupling, $U=0.1J$. Here, we see only minor deviations of the SPA-ME result from CI. The differences  become more pronounced for $T=12.5\hbar/J$ for oscillations with large amplitude, i.e. for $\tau \sim 20\hbar/J$. Further, we observe in the course of the relaxation process that there is a change in the frequencies of the oscillations. Additionally, as the propagation time increases, there is an increase in the amplitudes of the oscillations, with this trend continuing until $T=12.5\hbar /J$ and then being reversed. 

For $U=0.4J$ we observe the previously mentioned deviations of the SPA-ME result from the CI data. Only for $T =5\hbar/J$ good agreement is visible. Already for $T=10\hbar/J$ this holds only up to a relative time of $\tau=5\hbar/J$. Afterwards we observe that there is a shift of the phase of the oscillations, where significant deviations in the amplitude increase. In particular, while the exact result shows a damping of the oscillatory behavior, the SPA-ME shows the opposite trend. This behavior is also observed in a stronger form for later center-of-mass times.  It is noticeable that the minima of the density response function are shifted downward for the SPA-ME with increasing center-of-mass time, while there is no significant increase in the maxima. 
\begin{figure}[h]
    \centering
    \includegraphics[width=\columnwidth]{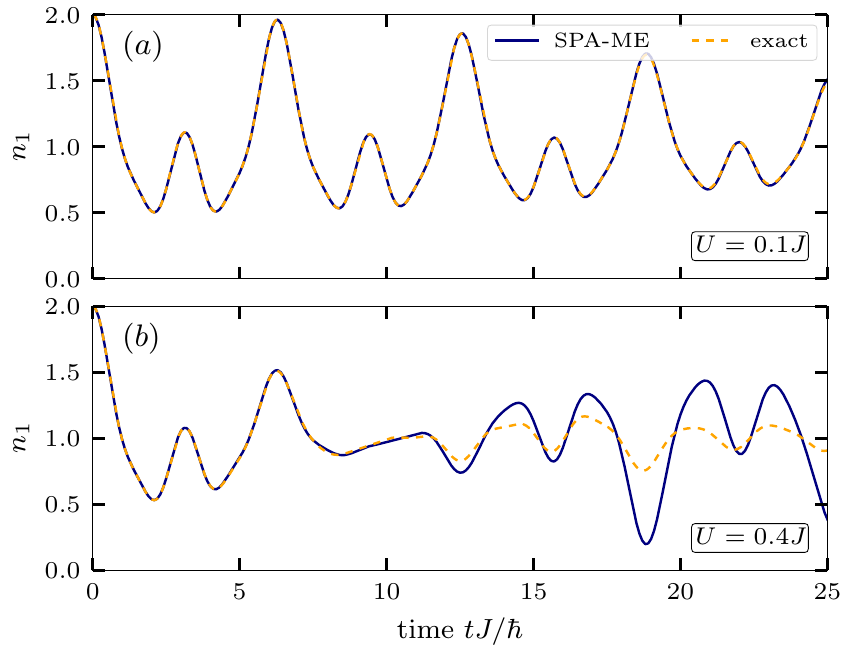}
    \caption{Density evolution at site 1 of a half-filled ring with 6 sites following a confinement quench.}
    \label{fig:density_evolution_ns6}
\end{figure}
The observed damped behavior of the exact dynamics of the density response function at $U=0.4J$ is also found for the density evolution shown in Fig.~\ref{fig:density_evolution_ns6} (b). Here, we again observe that SPA-ME overestimates the oscillations while still reproducing the exact phase. In (a) we see excellent agreement of the SPA-ME and exact results similar to the observed agreement for the density response function. \\

\begin{figure*}
    \centering
    \includegraphics[width=2\columnwidth]{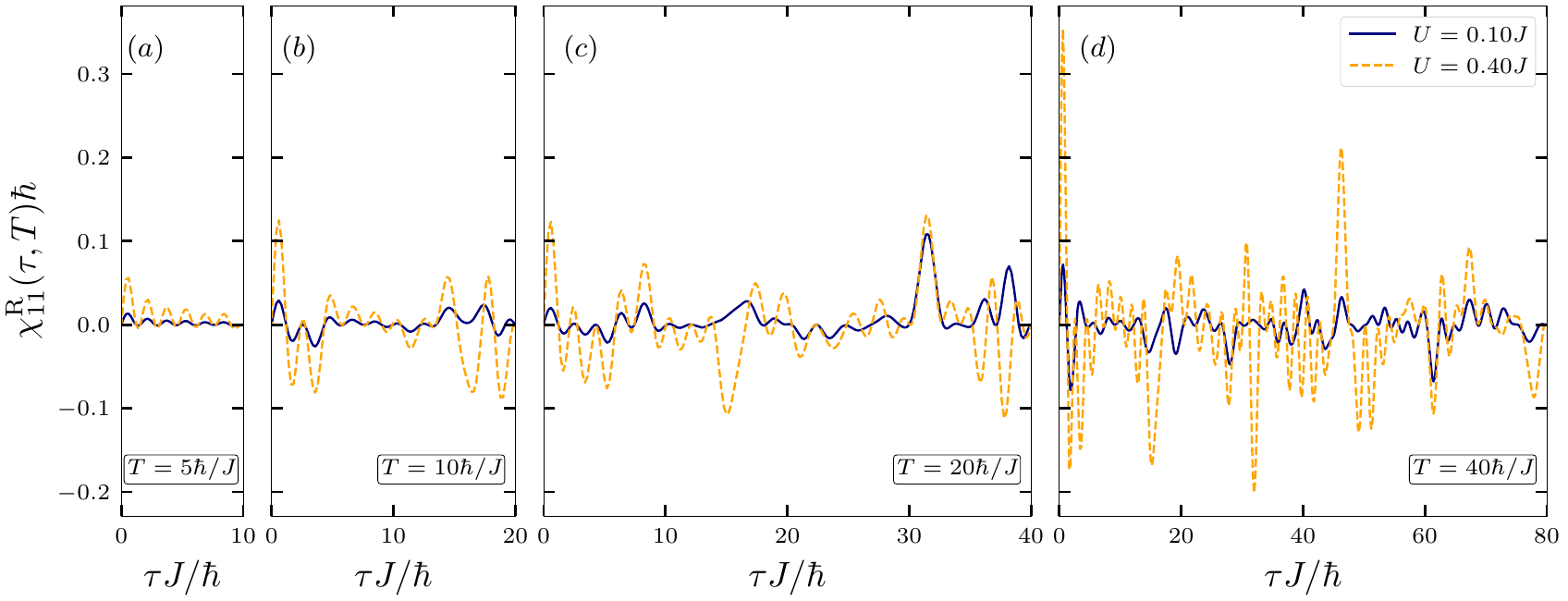}
    \caption{Density response function for different center-of-mass times $T$ following a confinement quench, for a half-filled ring with 30 sites  at $U=0.1J$ and $U=0.4J$ using SPA-ME.}
    \label{fig:neq_drf_ns30}
\end{figure*}

Finally, we turn again to larger systems for which, however, no CI data are available for comparison. Specifically, we consider a half-filled ring with 30 sites at $U=0.1J$ and $U=0.4J$, respectively, following an analogous confinement quench. Figure~\ref{fig:neq_drf_ns30} displays the density response function versus relative time, for fixed center-of-mass times. Here, we see for $T=5 \hbar/J$ that the oscillations of the density response function are damped with increasing relative time. Moreover, we find that the phases of the oscillations are the same for both on-site interaction strengths, which is also observable for all later center-of-mass times with only minor deviations between the two. Again, we find that stronger coupling leads to an increase of the amplitude. However, there is no damping of the oscillations for either coupling strength for larger center-of-mass times. \\

\begin{figure}[h]
    \centering
    \includegraphics[width=\columnwidth]{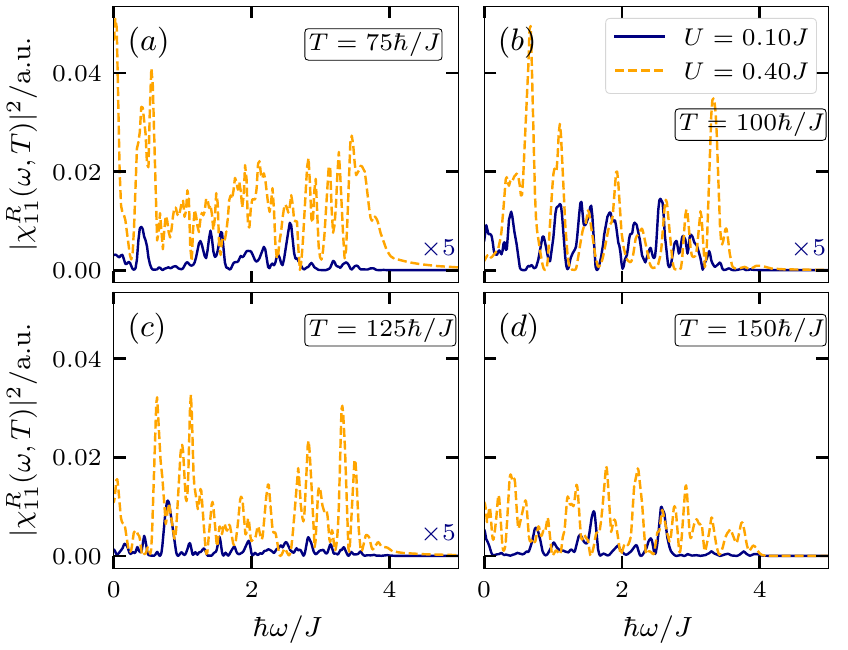}
    \caption{Fourier transform of the density response function with respect to the relative time for a half-filled ring with 30 sites at fixed center-of-mass times, $T$, at $U=0.1 J$ and $U=0.4J$ using SPA-ME. The $U=0.1J$ results in (a), (b) and (c) have been multiplied with a factor of $5$. A damping constant of $\eta = 0.04 J/\hbar$ was used.}
    \label{fig:neq_spec_ns30}
\end{figure}
\begin{figure}
    \centering
    \includegraphics[width=\columnwidth]{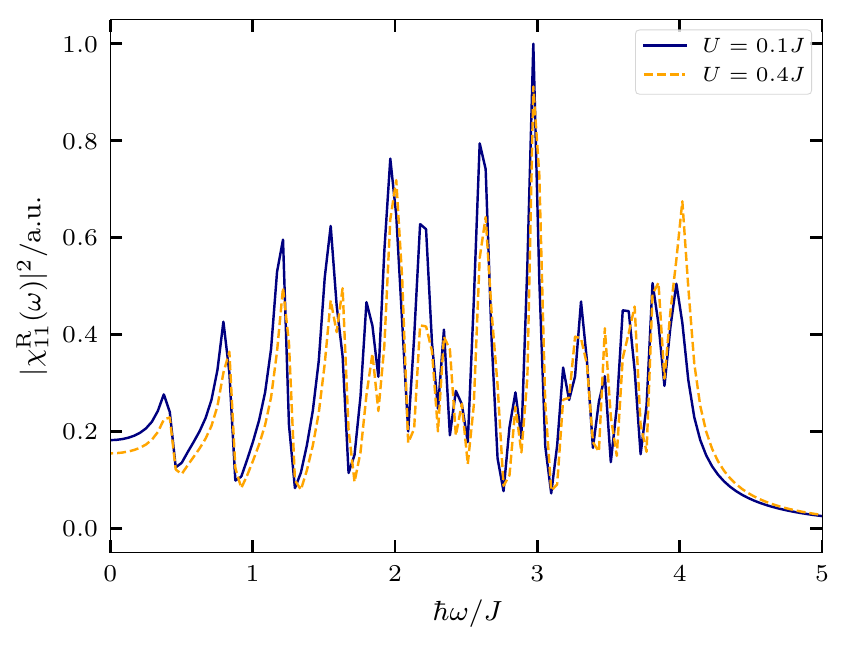}
    \caption{Same as Fig.~\ref{fig:neq_spec_ns30}, but for the ground state at $U=0.1J$ and $U=0.4J$.}
    \label{fig:eq_spec_ns30}
\end{figure}
Lastly, we consider nonequilibrium density response spectra of the same system at $U=0.1J$ and $U=0.4J$, for large center of mass times, $T\ge 75$, which are shown in Fig.~\ref{fig:neq_spec_ns30}. These nonequilibrium spectra are obtained via Fourier transform of the density response function with respect to the relative time $\tau$, for fixed center-of-mass times $T$. Here, we see a relatively similar structure of the spectra for both on-site interactions for $T=100\hbar/J$. However, the transition probabilities are significantly larger for $U=0.4J$ than for $U=0.1J$, for all center-of-mass times. Additionally, we find that the transition probabilities for excitations are largest for $T=150\hbar/J$ at $U=0.1J$, compared to the other center-of-mass times, whereas this is the case for $T=100\hbar/J$ at $U=0.4J$. 

Finally, we inquire whether the nonequilibrium spectra, for later times $T$, begin to approach the ground state spectra. To this end, we have computed ground state results for the same system that are displayed in Fig.~\ref{fig:eq_spec_ns30}. A first observation is that the nonequilibrium transition probabilities are significantly smaller than the ground state counterparts. Moreover, we find that the overall structure of the nonequilibrium spectra for $T=150\hbar/J$ starts to resemble the ground state spectra, at least, for $U=0.4J$. Similarly to the ground state results, the nonequilibrium spectra also vanish for frequencies larger than $4 J/\hbar$. Clearly, for a better quantitative comparison, longer nonequilibrium simulations would be desirable, as well as a comparison to equilibrium spectra at a finite temperature that takes into account the quench energy. This analysis is subject of ongoing work.

\section{Discussion and Outlook}\label{s:discussion}
In this paper we have extended our recent work on a novel quantum fluctuations approach presented in Ref.~\cite{schroedter_cmp_22}. This approach starts from single time single-particle density matrix operators and investigates their fluctuations around their expectation value -- the single-particle density matrix (or time-diagonal Green function). The approach is conceptionally similar to the classical fluctuations approach of Klimontovich \cite{klimontovich_jetp_57,klimontovich_1982}  as well as the stochastic mean field approximation of Ayik, Lacroix and others \cite{ayik_plb_08,lacroix_epj_14,lacroix_prb14}. 
All these methods have the attractive feature that correlation effects are mapped onto the dynamics of an ensemble of fluctuating single-particle quantities, and quantum expectation values are replaced by semi-classical mean values.

The advantage of the present approach is that it starts directly from the equations of motion for the nonequilibrium Green functions (NEGF) which allows to use well-known many-body approximations for the selfenergy, for reference, e.g. \cite{stefanucci-book,balzer-book,schluenzen_jpcm_19}. This also allows us to use results from recent dramatic accelerations of NEGF simulations that were achieved by using the Hartree-Fock GKBA and its transformation to time-local equations -- the G1--G2 scheme \cite{schluenzen_cpp16, schluenzen_prl_20,joost_prb_20,joost_prb_22}. There the gain in computation speed comes at the price of storing the nonequilibrium two-particle Green function $\mathcal{G}_2$ the size of which rapidly increases with the system size. One way to reduce the basis dimension is an embedding approach \cite{balzer_prb_23}. On the other hand, the present quantum fluctuations approach is even more powerful as it avoids the computation of $\mathcal{G}_2$ in favor of single-particle fluctuations $\delta \hat G_{ij}(t)$ and their classical counterparts $\Delta G^\lambda_{ij}(t)$. This method was shown to accurately reproduce G1--G2 simulations for single-time quantities, within the nonequilibrium $GW$ approximation \cite{schroedter_cmp_22}. In the present paper, we presented a major further extension to two-time observables that are composed of single-particle fluctuations: the general correlation function $L_{ijkl}(t,t')$, the density correlation function $\chi^R_{ij}(t,t')$ and the dynamic structure factor, $S(q,\omega)$.

To achieve this goal we first developed the multiple ensembles idea which allows one to compute commutators within the present semi-classical approach. This has led to the multiple ensembles stochastic polarization (SPA-ME) approximation that was extensively tested for Fermi-Hubbard chains in the numerics part of the paper. The first tests were devoted to the density response in the ground state. There the, SPA-ME approach is conceptionally analogous to the $GW$ approximation of Bethe-Salpeter theory and the random phase approximation of linear response theory \cite{kwong_prl_00}. We demonstrated excellent agreement with the RPA within the validity range of the approximation, i.e. for $U/J \lesssim 1$.

The main advantage of the present approach is its general applicability to correlated quantum systems in nonequilibrium. We demonstrated this for two types of external excitation: the first was a rapid potential kick applied to one site of the system. The second was a confinement quench in a system that was initially doubly occupied in its central region and then expands towards half filling. In both cases the SPA-ME approach allowed us to study the  buildup of the nonequilibrium density response $\chi^R(\tau,T)$ as a function of physical time $T$ that as passed after the excitation. Performing benchmarks against CI simulations, for small systems, we concluded that, for weak coupling, the method is very accurate. With increasing coupling, only the main features of the nonequilibrium spectra are captured, as expected from a weak coupling approximation. We then turned to larger systems for which our method is expected to be more accurate \cite{schroedter_cmp_22}. 

Based on these first proof of principle results we conclude that the present SPA-ME approach can be successfully applied to large systems for which the nonequilibrium density response and the impact of correlation effects can be studied for long propagation times. Work on a more systematic analysis of the nonequilibrium dynamics, both linear and nonlinear, and its dependence on the coupling strength and excitation scenario is presently in progress.
A particularly interesting excitation scenario is to use a short spatially monochromatic pulse.
As was shown in Ref.~\cite{kwong_prl_00}, nonequilibrium $GW$ simulations of a uniform system following such an excitation are expected to yield high quality dynamic structure factors that are sum rule preserving and include vertex corrections, thereby going well beyond RPA.

\section*{Acknowledgements}
We acknowledge stimulating discussions with Christopher Makait.
This work is supported by the Deutsche Forschungsgemeinschaft via project BO1366/16 and through high-performance computing resources available at the Kiel University Computing Centre.



\appendix
\section{Analytical result for the density response function of noninteracting Hubbard clusters}\label{s:appendix}


The hamiltonian of a 1D tight-binding chain with perio\-dic boundary conditions is given by
	\begin{equation}
		\hat{H}_0=-J\sum_{\braket{n,m}}\sum_{\sigma}\hat{c}_{n\sigma}^\dagger \hat{c}_{m\sigma}\,,
		\label{eq:h0-tight-binding}
	\end{equation}
where only chains with \(4\mathbb{N}-2\) sites will be considered. By transforming into the momentum basis
	\begin{align}
		\hat{c}_{n\sigma}^\dagger=&\frac{1}{\sqrt{N}}\sum_{q}\mathrm{e}^{iqm}\hat{c}_{q\sigma}^\dagger\,,\\
        q=&\frac{\pi}{N}\cdot\left(-N\,,-N-2\,,\cdots\,,N-2\right)\,,
	\end{align}
	Eq.~\eqref{eq:h0-tight-binding} takes the form
	\begin{equation}
		\hat{H}_0=-2J\sum_{\sigma,q}\cos(q)\,\hat{c}_{q\sigma}^\dagger \hat{c}_{q\sigma}\,,
	\end{equation}
where the time-dependent version of the field operators is given by
	\begin{equation}
		\hat{c}_{q\sigma}(t)
		=\mathrm{e}^{-\frac{2iJ}{\hbar}\cos(q) t}\,\hat{c}_{q\sigma}\,.
	\end{equation}
	From this we obtain the diagonal elements of the density operator for site $m$,
	\begin{equation}
		\hat{n}_{m\sigma}(t)=\hat{c}_{m\sigma}^\dagger(t)\hat{c}_{m\sigma}(t)=\frac{1}{N}\sum_{q,q^\prime}\mathrm{e}^{i(q-q^\prime)m}\hat{c}_{q\sigma}^\dagger(t)\hat{c}_{q^\prime\sigma}(t)\,,
	\end{equation}
	and the commutator
	\begin{align}
		\left[n_{m\sigma}(t),n_{n\sigma^\prime}\right]=&\frac{2i}{N^2}\delta_{\sigma,\sigma^\prime}\sum_{q,k,k^\prime}\sin\Big\{(q-k)m\\
        &+(k-k^\prime)n-\frac{2J}{\hbar}\left[\cos(q)-\cos(k)\right]t\Big\}\\
        &\hat{c}_{q\sigma}^\dagger \hat{c}_{k^{\prime}\sigma}\,.
	\end{align}
	Thus, for zero temperature and half filling we obtain for the density response function
	\begin{align}
		\chi^R_{mn}(t)=&\Theta(t)\frac{4\hbar}{N^2}\sum_{q,k}\sin\Big\{(q-k)(m-n )\\
        &-\frac{2J}{\hbar}\left[\cos(k)-\cos(q)\right]t\Big\}\Theta[\cos(q)]\,.
	\end{align}
	

\section{Random phase approximation for the Hubbard-model}\label{s:appendix_rpa}
For the Hubbard-Hamiltonian in one dimension with \(N\) lattice sites and 
	\begin{align}
		\hat K=\hat H-\mu \hat N=&\sum_{k\sigma}\epsilon_k \hat c_{k\sigma}^\dagger \hat c_{k \sigma}\\
        &+\frac{U}{2N}\sum_{\sigma}\sum_{k,k^\prime,q} \hat c_{k+q\sigma}^\dagger \hat c_{k^\prime-q-\sigma}^\dagger \hat c_{k^\prime-\sigma} \hat c_{k\sigma}\,,
	\end{align}
	where \(\epsilon_k=-2J\cos(k)-\mu\), the density response function in RPA takes the form
	\begin{equation}
		\chi_{\tn{RPA}}^{\tn{R}}(q,\omega) = \frac{2\Pi^{\tn{R}}(q,\omega)}{1-U\Pi^{\tn{R}}(q,\omega)}\,,\label{eq:RPA_density_density_responsefunction}
	\end{equation}
	where
	\begin{equation}
		\Pi^{\tn{R}}(q,\omega)=\frac{1}{N}\sum_{k}\frac{n(\epsilon_k)-n(\epsilon_{k+q})}{\hbar\omega+\epsilon_{k}-\epsilon_{k+q}+i\eta}\,,
  \label{eq:RPA_polarization_function}
	\end{equation}
	and \(n(\epsilon_q)=\Theta[\cos(q)]\), for zero temperature and half filling. The missing factor of two in Eq. \eqref{eq:RPA_polarization_function} and the extra factor of two in Eq. \eqref{eq:RPA_density_density_responsefunction} in comparison to the conventional form of these functions (compare \cite{altland_simons_2010}) are due to the unusual spin-indices in the Hamiltonian. 
 


%

\end{document}